\documentclass[seceq]{ptptex}

\usepackage{graphicx}

\renewcommand{\Re}{\mathop{\rm Re}}
\newcommand{\Tr}{\mathop{\rm Tr}}
\newcommand{\ch}{{\rm ch}}

\title{Numerical study of staggered quark action on quenched \\
  anisotropic lattices}

\author{
Kouji \textsc{Nomura}$^{1,}$%
  \footnote{E-mail: nomura1@hiroshima-u.ac.jp},
Hideo \textsc{Matsufuru}$^{2,}$%
  \footnote{E-mail: hideo.matsufuru@kek.jp},
and Takashi \textsc{Umeda}$^{3,}$%
  \footnote{E-mail: tumeda@yukawa.kyoto-u.ac.jp},
}

\inst{
$^1$Department of Physics, Hiroshima University,\\
Higashi-hiroshima 739-8626, Japan\\
$^2$Computing Research Center,
High Energy Accelerator Research Organization (KEK),
Tsukuba 305-0801, Japan \\
$^3$Yukawa Institute for Theoretical Physics, Kyoto University,\\
Kyoto 606-8502, Japan
}

\abst{
The staggered quark action on anisotropic lattices is studied.
We carry out numerical simulations in the quenched approximation at 
three values of lattice spacing ($a_{\sigma}^{-1}=1-2$ GeV) with the
anisotropy $\xi= a_{\sigma}/a_{\tau}=4$, where $a_{\sigma}$
and $a_{\tau}$ are the spatial and temporal lattice spacings,
respectively. 
The bare anisotropy $\gamma_F$ in the quark action is
numerically tuned through the ratio of meson masses in the fine and coarse
directions, and through the dispersion relation of a meson,
so that the renormalized fermionic anisotropy coincides with that of
the gauge field.
The discrepancy between these two calibration schemes provides
an estimate of the finite lattice artifact, which is found to be
sizable in the range of cutoff explored in this work.
We also compute the meson masses using correlators with the wall
source at the tuned anisotropy parameter.
The flavor symmetry breaking effect smoothly decreases as
$\beta$ increases.
The effect of uncertainty in $\gamma_F$ on the meson masses
are examined.
We also discuss a perspective on dynamical simulations.}

\begin{document}

\maketitle


\section{Introduction}

In numerical studies of lattice QCD, one frequently encounters
a case which requires a fine lattice spacing in the temporal
direction while does not comparatively in the spatial directions.
Increasing the lattice cutoffs in all the four directions
severely increases the computational cost,
since it increases at least in proportion to the volume of
the lattice, and in fact more rapidly in particular in dynamical
simulations \cite{Bernard02}.
A solution is provided by anisotropic lattices \cite{Karsch82},
on which the temporal lattice spacing is finer than the spatial ones.
The technique is useful in various fields of the lattice QCD simulation:
At finite temperature, a large number of the degrees of freedom
in the Euclidean time direction leads a large number of Matsubara
frequencies, which is efficient for calculations of the equation of
state \cite{EOS} and for analyses of temporal correlation functions
of hadrons \cite{TARO01,Umeda01,Ishii02,Umeda03,Asakawa03}.
The large temporal cutoff is important for
relativistic formulations of heavy quark on the lattice
\cite{Klassen99,Chen01,heavy1,Okamoto02,heavy2}. 
It is also efficient when the signal-to-noise ratio deteriorate
quickly, as in the cases of glueballs \cite{Morningstar99,Ishii02},
negative parity baryons \cite{Nemoto03},
and the pion scattering length \cite{Liu02}.

On the other hand, on anisotropic lattices one has additional
parameters in the actions which control the anisotropies of the fields.
In general these anisotropy parameters should be tuned numerically.
Inappropriate tuning of the parameters breaks the rotational symmetry
of the lattice, and may lead unphysical results.
The uncertainties in the tuning of the parameters bring additional
errors into observed quantities.
For precise calculations, one needs to tune anisotropy parameters
with good statistical accuracy and to control the systematic
errors in the continuum extrapolation.
In this paper, we investigate this calibration process in detail for
the staggered quark action.

To explain the situation we are faced with in more detail,
let us focus on a study of hadrons at finite temperature,
which is a main motivation of present work.
To investigate the hadron structure at $T>0$, one needs to
treat the hadron correlators in the temporal direction
\cite{HNS93,TARO01,Umeda01}.
Because the lattice size in the temporal direction
is limited to $N_t a_\tau=1/T$, where $T$ is the temperature,
near and above the critical temperature the number of degrees of freedom
is severely limited without introducing the anisotropic lattice.
For example, the number of degrees of freedom is significant for
reliable extraction of the spectral function from
the lattice data \cite{NAH99,Umeda03,Asakawa03}.

So far the studies of hadrons at finite temperature have been
performed mainly on quenched lattices.
Dynamical simulations are manifestly important,
as exhibited by the fact that the order of phase transition
changes as the number of dynamical quark flavors varies.
As the quark action we adopt the staggered action,
which has several advantages over the Wilson-type quarks \cite{CPPACS}:
Firstly, the staggered action retains the remnant of the chiral symmetry.
This is important to investigate the role of the chiral
symmetry near the phase transition.
Secondly, one can explore smaller quark mass region than the cases
with the Wilson-type formulations.
Thirdly, the cost of numerical simulation is much more economical
than the other formulations.
A disadvantage lies in its complication in the flavor structure.
To circumvent the effects of the flavor symmetry breaking,
improved versions of the staggered fermion have been developed.

In this work, however, we adopt the simplest version
without any improvement.
This is because an improvement adds the anisotropy parameters
which is to be tuned in general nonperturbatively.
At this first stage of development of the anisotropic staggered action,
we concentrate on the most significant effect of the anisotropy
on the spectrum.
For the gauge field we adopt the standard Wilson plaquette action,
which has the discretization errors in the same order as the quark
action we adopt.
The same combination of actions have also been used
in Ref.\cite{DynamicalAniso} for a dynamical simulation,
while their discussion on the systematic errors and statistical
precision were not sufficient for our present purposes.

The dynamical anisotropic lattices are more involved than the quenched
case, since one needs to tune the anisotropy parameters for the gauge
and quark fields simultaneously.
So far there is no systematic investigation of anisotropic staggered
quarks even in the quenched approximation.
In this paper we therefore concentrate on the quenched approximation
and investigate the properties of the staggered quark on anisotropic
lattices.

The numerical simulations are performed on quenched lattices with
three lattice spacings, at fixed renormalized anisotropy $\xi=4$. 
These scales cover the range of the spatial lattice cutoff
$a_{\sigma}^{-1}=1-2$ GeV.
We apply two calibration procedures with different definitions of
fermionic anisotropy $\xi_F$: with the ratio of the masses in
the fine and coarse directions, and through the meson dispersion
relation.
The former has an advantage in statistical precision, while the
latter can be used even in the heavy quark mass region.
Differences between the results of these two procedures signal
the finite lattice artifacts.
For the reason of statistical fluctuations, we use only the
pseudoscalar channel in the calibration.
Other mesonic channels are observed in the second part of the paper
by employing the wall source, and compared with the results on the
isotropic lattices.

This paper is organized as follows.
The next section briefly summarizes the anisotropic staggered action.
Section~\ref{sec:calibration_procedures} describes the calibration
procedures.
The numerical result of the calibration is presented
in Sec.~\ref{sec:calibration}.
Section~\ref{sec:spectroscopy} describes the spectroscopy with the wall
source.
Section~\ref{sec:conclusion} is devoted to our conclusions
in the quenched simulations.
In the last section, we give our perspective toward the dynamical
simulations with the staggered quarks.


\section{Staggered quark on anisotropic lattices}
 \label{sec:action}

\subsection{Anisotropic lattice actions}

The gauge field is described with the standard Wilson gauge action,
\begin{equation}
S_G  =  \beta \sum_x \left\{ \sum_{i<j=1}^{3} \frac{1}{\gamma_G}
 \left[ 1-\frac{1}{3}\Re \Tr  U_{ij}(x) \right] + \right.
        \left.  \sum_{i=1}^{3} \gamma_G
 \left[ 1-\frac{1}{3}\Re \Tr  U_{i4}(x) \right] \right\} ,
\end{equation}
where $\beta=6/g_0^2$ is the bare coupling and
$\gamma_G$ the bare gauge anisotropy parameter.
The plaquette variable $U_{\mu\nu}(x)$ is defined with the link
variable $U_\mu(x)\simeq e^{iga_\mu A_\mu (x)} \in$SU(3) as
\begin{equation}
 U_{\mu\nu}(x) =
   U_\mu(x) U_\nu(x+\hat{\mu}) U^\dag_\mu(x+\hat{\nu}) U^\dag_\nu(x).
\end{equation}
A lattice site is labeled by an integer vector $x$ whose component
$x_\mu$ is in units of lattice spacing $a_\mu$,
where $a_1=a_2=a_3=a_\sigma$ and $a_4=a_\tau$.
$\hat{\mu}$ is a unit vector in the $\mu$-th direction.

The staggered quark action on an anisotropic lattice is defined as
\begin{eqnarray}
  S_F  =  \sum_{x,y} \bar{\chi}(x) K(x,y) \chi(y),
\end{eqnarray}
\begin{eqnarray}
  K(x,y)  &=& \delta_{x,y}
   - \kappa_\sigma \sum_{i=1}^3
   \eta_i(x) \left[ U_i(x)\delta_{x+\hat{i},y}-
              U^\dagger_i(x-\hat{i})
                           \delta_{x-\hat{i},y}\right] \nonumber\\
& &\hspace{8mm}
    - \gamma_F\kappa_\sigma
        \eta_4(x)\left[  U_4(x)\delta_{x+\hat{4},y}-
               U^\dagger_4(x-\hat{4})
                           \delta_{x-\hat{4},y}\right],
\end{eqnarray}
where  $\gamma_F$ is the bare anisotropy of the quark field,
$\kappa_{\sigma}= 1/2m_q$ with $m_q$ the bare quark mass in spatial
lattice units.
$\eta_\mu(x)$ is the staggered phase,
$\eta_\mu(x)=(-1)^{x_1+\cdots +x_{\mu-1}}$, with
$\eta_1(x)=1$.

The staggered quark fields $\chi$ and $\bar{\chi}$ have only
the color components.
The 4-spinor field with four degenerate flavors, $\psi_\alpha^f(X)$,
where $X$ labels the space-time position in units of $2a_\mu$,
are represented as a linear combination of $\chi(x)$ as
\cite{Rothe,KSMNP83}
\begin{eqnarray}
  \psi_\alpha^f(X) &=& \frac{1}{\sqrt{2}}
          \sum_\rho  (T_\rho)_{\alpha f} \chi_\rho (X), \\
  T_\rho &=& \gamma_1^{\rho_1} \gamma_2^{\rho_2}
             \gamma_3^{\rho_3} \gamma_4^{\rho_4},  \\
  \chi_\rho (X) &=& \chi (2X+\rho),
 \label{eq:chi_of_X}
\end{eqnarray}
where $\rho$ labels a position in the hypercube and its components
are 0 or 1.
Therefore the quantum number of a quark bilinear is specified
by its spin-flavor structure \cite{KSMNP83,Golterman86}.

\subsection{Meson correlators}

The meson correlators are composed of the quark propagator,
\begin{equation}
   S(x,y) = \langle \chi(x) \bar{\chi}(y) \rangle  = K^{-1}(x,y),
\label{eq:q_prop}
\end{equation}
which is obtained by solving a linear equation
\begin{equation}
  \sum_x K(z,x) S(x,0) = b(\vec{z})\delta_{z_4,0} ,
\end{equation}
where the vector $b(\vec{z})$ is a source field.
(The color index is omitted.)
As $b(\vec{z})$, we select the following two cases;
(i) the point source, $b_p(\vec{y})=\delta_{\vec{y},0}$, and
(ii) the wall sources, $b_e(\vec{y})=1$ and
 $b_o(\vec{y})=(-1)^{y_1+y_2+y_3}$.
In the latter case, the gauge must be fixed.

In this paper, we treat single time-slice correlators,
namely the sink operators at each $t$ are defined only in a spatial
plane on a single time slice \cite{GSST84}.
Such a correlator contains two modes, one
monotonously and the other oscillatingly decay.
The latter has the opposite parity to the former.
In the following, the correlators are labeled by their
non-oscillating channels.

The meson correlators composed of the quark propagator with the
point source are used in the calibration (Sec.~\ref{sec:calibration}).
Since we do not fix the gauge at this stage, the correlators must be
constructed with the gauge invariant ingredients.
The pseudoscalar and vector correlators are represented as
\begin{eqnarray}
   M_{PS}(t) &=& \sum_{\vec{x}}  |S_p(x)|^2 ,
\label{eq:meson_point1} \\
   M_{V_j} (t) &=& \sum_{\vec{x}} (-)^{x_j} |S_p(x)|^2
 \hspace{0.5cm} (j=1,2,3),
\label{eq:meson_point2}
\end{eqnarray}
where $S_p(x)$ is the quark propagator with the point source
at the origin.
These correlators have spin-flavor structures
$\gamma_5 \otimes \gamma_5$ and $\gamma_j \otimes \gamma_j$,
respectively.
While these representations are for the vanishing meson momentum,
the momentum insertion is straightforward.
In the case of vector correlator, the finite momentum state needs
some care, since the operator mixes with the fourth
component.
However we do not discuss this problem further, since we decide
to use only the pseudoscalar correlator in the calibration
because of large fluctuation in the vector channel.
We also measure the correlators in the $z$-direction to define
the anisotropy through the ratio of masses in the fine and coarse
directions.
Rewriting Eqs.~(\ref{eq:meson_point1}) and (\ref{eq:meson_point2})
for the correlators in $z$-direction is also straightforward.

The correlators composed of the quark propagators with
wall sources are as follows \cite{GGKS91}.
\begin{eqnarray}
   M_{\pi}(t) &=& \sum_{\vec{x}}
   \left[ S_e^\dag(x)S_e(x) + S_o^\dag(x)S_o(x) \right]
\label{eq:meson_wall1}
\\
   M_{\tilde{\pi}}(t) &=& \sum_{\vec{x}} (-)^{x_1+x_2+x_3}
   \left[ S_e^\dag(x)S_o(x) + S_o^\dag(x)S_e(x) \right]
\label{eq:meson_wall2}
\\
   M_{\pi_3}(t) &=& \sum_{\vec{x}}
   \left[   S_e^\dag(x+\hat{3})S_e(x)
          - S_o^\dag(x+\hat{3})S_o(x) \right]
\label{eq:meson_wall3}
\\
   M_{\tilde{\pi}_3}(t) &=& \sum_{\vec{x}} (-)^{x_1+x_2+x_3}
   \left[   S_e^\dag(x+\hat{3})S_o(x)
          - S_o^\dag(x+\hat{3})S_e(x) \right]
\label{eq:meson_wall4}
\\
   M_{\rho_6^A}(t) &=& \sum_{\vec{x}}
   \left[   S_e^\dag(x+\hat{2}+\hat{3})S_e(x)
          + S_o^\dag(x+\hat{2}+\hat{3})S_o(x) \right]
\label{eq:meson_wall5}
\\
   M_{\rho_6^B}(t) &=& \sum_{\vec{x}} (-)^{x_1+x_2+x_3}
   \left[   S_e^\dag(x+\hat{2}+\hat{3})S_o(x)
          + S_o^\dag(x+\hat{2}+\hat{3})S_e(x) \right]
  \hspace{0.5cm}
\label{eq:meson_wall6}
\end{eqnarray}
The spin-flavor structures of these correlators are;
$\pi$:            $\gamma_5 \otimes \gamma_5$,
$\tilde{\pi}$:    $\gamma_4\gamma_5 \otimes \gamma_4\gamma_5$,
$\pi_3$:          $\gamma_5 \otimes \gamma_5\gamma_3$,
$\tilde{\pi}_3$:  $\gamma_4\gamma_5 \otimes \gamma_4\gamma_5\gamma_3$,
$\rho_6^A$:       $\gamma_3 \otimes \gamma_2$, and
$\rho_6^B$:       $\gamma_3\gamma_4 \otimes \gamma_2\gamma_4$.
In addition to these quantum numbers, each correlator contains the
opposite parity channel as the oscillating mode.
The correlator (\ref{eq:meson_wall1}) has the same quantum number as
Eq.~(\ref{eq:meson_point1}).
On the other hand, the quantum number of the correlators
(\ref{eq:meson_wall5}) and (\ref{eq:meson_wall6}) are different
from that of Eq.~(\ref{eq:meson_point2}) at finite lattice spacing.

At large $t$,
the correlator approaches a form \cite{GSST84}
\begin{equation}
   M(t) \rightarrow Z \exp(-mt) + (-)^t \tilde{Z} \exp(-\tilde{m}t).
\label{eq:fit_form}
\end{equation}
To extract the meson masses,
we fit the numerical data of the meson correlator with the wall source
to this form after adding the contribution via the temporal boundary.
In the calibration, only the first term in Eq.~(\ref{eq:fit_form})
is retained since the contribution from the second term quickly
disappears in the pseudoscalar channel.


\section{Calibration procedures}
 \label{sec:calibration_procedures}

\subsection{Calibration schemes}

The anisotropy parameters $\gamma_G$ and $\gamma_F$ should be tuned
so that the physical isotropy condition,
$\xi_G(\gamma_G^*,\gamma_F^*)=\xi_F(\gamma_G^*,\gamma_F^*)=\xi$,
holds at each set of $\beta$ and $m_q$,
where $\xi_G(\gamma_G,\gamma_F)$ and
$\xi_F(\gamma_G,\gamma_F)$ are the renormalized anisotropies
defined through the gauge and fermionic observables, respectively.
In quenched simulations, one can firstly determine
$\gamma_G^*$ independently of $\gamma_F$, and then tune
$\gamma_F^*$ for fixed $\gamma_G^*$.
In contrast, $\gamma_G$ and $\gamma_F$ need to be tuned
simultaneously in dynamical simulations.

In this work, we use the values of the gauge parameters $\beta$ and
$\gamma_G$ whose renormalized anisotropy, $\xi_G$, has been
determined in good accuracy.
The values of the parameters used will be described in detail
in Sec.~\ref{subsec:lattice_setup}.
For the quark field, we employ the following two calibration
schemes accordingly to the definitions of fermionic anisotropy
$\xi_F$.

\begin{itemize}

\item
{\it Mass ratio scheme}:
In this scheme,
$\xi_F$ is defined as the ratio of meson masses in the temporal and
spatial directions \cite{TARO01},
\begin{eqnarray}
 \xi_F^{(M)} \equiv m_{H}^{(z)}/m_{H}^{(t)} .
\end{eqnarray}
We use only the pseudoscalar channel as already noted in the previous
section for the statistical reason.

\item
{\it Dispersion relation scheme}:
An alternative definition of $\xi_F$ makes use of the
meson dispersion relation \cite{Klassen99},
\begin{equation}
 E^2(\vec{p})=m_H^2 +\vec{p}^{\,2}/(\xi_F^{(DR)})^2.
\label{eq:xi_F_DR}
\end{equation}
$E(\vec{p})$ and $m_H$ are in temporal lattice units and
$\vec{p}$ is in spatial lattice units, and hence $\xi_F$ appears.
On a finite volume lattice, $p_i=2\pi n_i/L_i$ ($i=1$,2,3),
where $L_i$ is the lattice size in the $i$-th direction.
Eq.~(\ref{eq:xi_F_DR}) requires that the rest mass equals
the kinetic mass.

\end{itemize}

\subsection{Free quark case}

The mass dependences of $\xi_F^{(M)}$ and $\xi_F^{(DR)}$
for the free quark propagator give us a guide to analyze
the results of the numerical simulation.
The free quark propagator of the staggered quark field
(\ref{eq:chi_of_X}) in the momentum space is \cite{Rothe}
\begin{equation}
  S_{\rho\rho'}(p) = \frac{
     -i\sum_j \Gamma_{\rho\rho'}^j(p)\sin(p_j/2)
     -i \Gamma_{\rho\rho'}^4(p) \gamma_F \sin(p_4/2)
    + m_q \delta_{\rho\rho'}}
  {  \sum_j \sin^2(p_j/2) + \gamma_F^2 \sin^2(p_4/2) + m_q^2 },
\end{equation}
where
\begin{equation}
    \Gamma_{\rho\rho'}^\mu(p) = e^{ip(\rho-\rho')/2}
   [\delta_{\rho+\hat{\mu},\rho'} + \delta_{\rho-\hat{\mu},\rho'} ]
   \eta_\mu (\rho) .
\end{equation}
The propagator in the $t$-direction,
\begin{equation}
  S_{\rho\rho'}(\vec{p},t) =
  \int_{-\pi}^{\pi} \frac{dp_4}{2\pi} e^{ip_4t} S(p),
\end{equation}
has asymptotic behavior
$S_{\rho\rho'}(\vec{p},t) \propto e^{-E^{(4)}(\vec{p})t}$,
with $E^{(4)}(\vec{p})$ implicitly given by
\begin{equation}
 \ch E^{(4)}(\vec{p})
 = 1 + \frac{2}{\gamma_F^2} \left[
               \sum_{j=1}^3 \sin^2\frac{p_j}{2} + m_q^2 \right].
\label{eq:Free_DR}
\end{equation}
Similarly, for the propagator in $z$-directions,
\begin{equation}
  \ch E^{(3)}(\vec{\tilde{p}})
  = 1 + 2 \left[ \sum_{j=1,2} \sin^2\frac{p_j}{2}
                 +  \gamma_F^2 \sin^2\frac{p_4}{2}
                 +  m_q^2  \right] ,
\end{equation}
where $\vec{\tilde{p}}=(p_1,p_2,p_4)$.
For a small quark mass, the rest masses,
$M_1\equiv E(\vec{p}=0)$, are represented as
\begin{eqnarray}
  M_1^{(4)} &=& \frac{2}{\gamma_F}m_q
              - \frac{1}{3\gamma_F^3}m_q^3 + O(m_q^4)
\nonumber \\
  M_1^{(3)} &=& 2 m_q - \frac{1}{3} m_q^3 + O(m_q^4).
\end{eqnarray}
Requiring $\xi_F^{(M)}\equiv M_1^{(3)}/M_1^{(4)}=\xi$,
\begin{equation}
  \gamma_F^{*(M)} = \xi \left[
    1 + \frac{1}{6}\left(1 - \frac{1}{\xi^2} \right) m_q^2
      + O(m_q^3).
\label{eq:calib_free_M}
\right]
\end{equation}
By differentiating Eq.~(\ref{eq:Free_DR}) by $p_i$,
the tuned anisotropy in the dispersion relation scheme is
obtained as
\begin{equation}
  \gamma_F^{*(DR)} = \xi \left[
    1 - \frac{1}{3\gamma_F^2} m_q^2 + O(m_q^4) \right].
\label{eq:calib_free_DR}
\end{equation}
Therefore in both the schemes the quark mass dependences
of $\gamma_F^*$ are of $O(m_q^2)$.
As the bare quark mass increases, $\gamma_F^{*(M)}$
increases while $\gamma_F^{*(DR)}$ decreases.
These results are useful to understand the results
of the numerical simulation.


\section{Numerical result of calibration}
 \label{sec:calibration}

\subsection{Lattice setup}
 \label{subsec:lattice_setup}

\begin{table}[tb]
\caption{
The lattice parameters.
The values of $\gamma_G$ are based on the results of Ref.
~\cite{Klassen98a} ($\beta=5.95$, 6.10) and 
Ref.~\cite{Matsufuru03} ($\beta=5.75$).
The lattice scale $a_{\sigma}^{-1}$ is set by the hadronic
radius $r_0$ \cite{Aniso01b,Matsufuru03}.
The mean-field values are from Ref.~\cite{Aniso01b}.
$^{*}$%
Note that those values at $\beta=5.75$ were evaluated at slightly
different anisotropy, $\gamma_G=3.072$.}
\begin{center}
\begin{tabular}{ccccccccc}
\hline \hline
$\beta$ & $\gamma_G$ & Size & $N_{conf}$ & $r_0$ & $a_{\sigma}^{-1}(GeV)$ &
$u_\sigma$ & $u_\tau$  \\
\hline
5.75 & 3.136~ & $12^2 \times 24 \times 96$~ & 224 & 2.786(15) & 1.100(6) &
   0.7620(2)$^*$ & 0.9871$^*$ \\
5.95 & 3.1586 & $16^2 \times 32 \times 128$ & 200 & 4.110(23) & 1.623(9) &
   0.7917(1) & 0.9891 \\
6.10 & 3.2108 & $20^2 \times 40 \times 160$ & 200 & 5.140(32) & 2.030(13)&
   0.8059(1) & 0.9901 \\
\hline\hline
\end{tabular}
\end{center}
\label{tab:parameters}
\end{table}

For the quenched anisotropic lattices, several results for
the calibration of gauge field are available
\cite{Burgers88,TARO97,KES98,Klassen98a,Matsufuru03}.
We adopt three sets of the lattice parameters ($\beta$,$\gamma_G$)
at the renormalized anisotropy $\xi=4$, according to
Refs.~\cite{Klassen98a,Matsufuru03}.
The parameters are summarized in Table~\ref{tab:parameters}.
These lattices have almost same physical volume.
The temporal direction is chosen as the fine direction,
and the size in the $z$-direction is set being the same physical
length as the $t$-direction.
The lattice scales are set by the hadronic radius $r_0$
\cite{Sommer94}.

For the larger two values of $\beta$, we adopt the Klassen's result
for $\gamma_G^*$ which was determined within 1\% statistical
accuracy \cite{Klassen98a}.
These two lattices are almost the same as those used
in Ref.~\cite{Aniso01b}, while the size in the $z$-direction
is doubled.

At $\beta=5.75$, we use a recent result of a precision calibration
using the hadronic radius $r_0$ \cite{Sommer94}
as a calibration condition \cite{Matsufuru03}.
With the L\"uscher-Weisz noise reduction technique \cite{LW01},
the static quark potential was computed to the level of 
0.2\% accuracy in the fine and coarse directions.
Since the value used in this paper was obtained in an earlier
stage of the work \cite{Matsufuru03}, $\gamma_G^*$ is slightly
different from the final result quoted in Ref.~\cite{Matsufuru03}.

We also quote the spatial and temporal mean-field values \cite{LM93},
$u_\sigma$ and $u_\tau$ respectively, in the Landau gauge determined in
Ref.~\cite{Aniso01b}.
For $\beta=5.75$, the values listed in Table~\ref{tab:parameters}
are not correct exactly, since they were determined at the different
$\gamma_G$ ($\gamma_G=3.072$).
This is not a serious problem, since we quote these values just
for a qualitative comparison of the numerical result of $\gamma_F^*$
with a mean-field estimate,
$\gamma_F^{*(MF)} = \xi\cdot (u_\tau /u_\sigma)$.

\subsection{Correlator data}

\begin{figure}[tb]
\includegraphics[width=7.0cm]{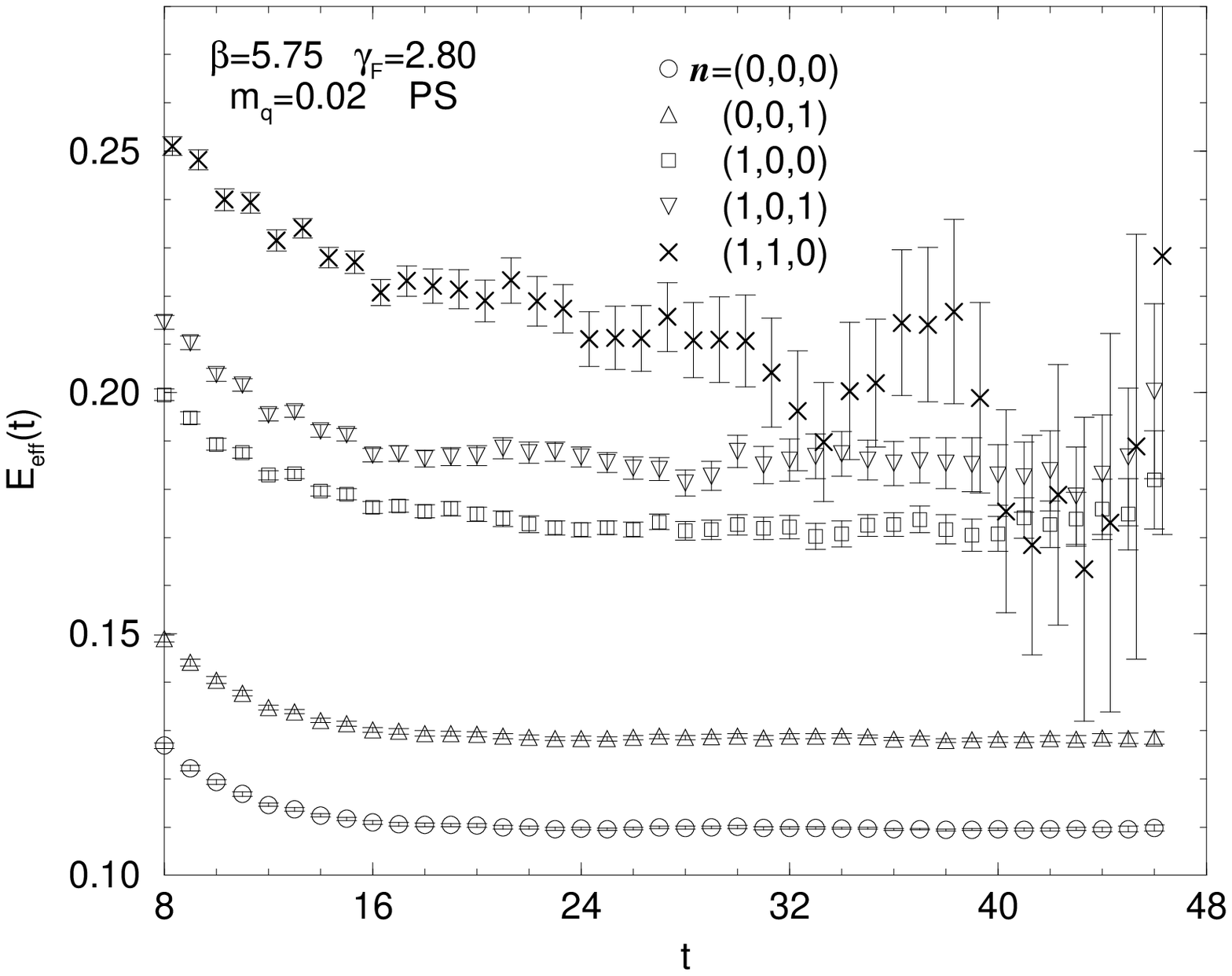}
\includegraphics[width=7.0cm]{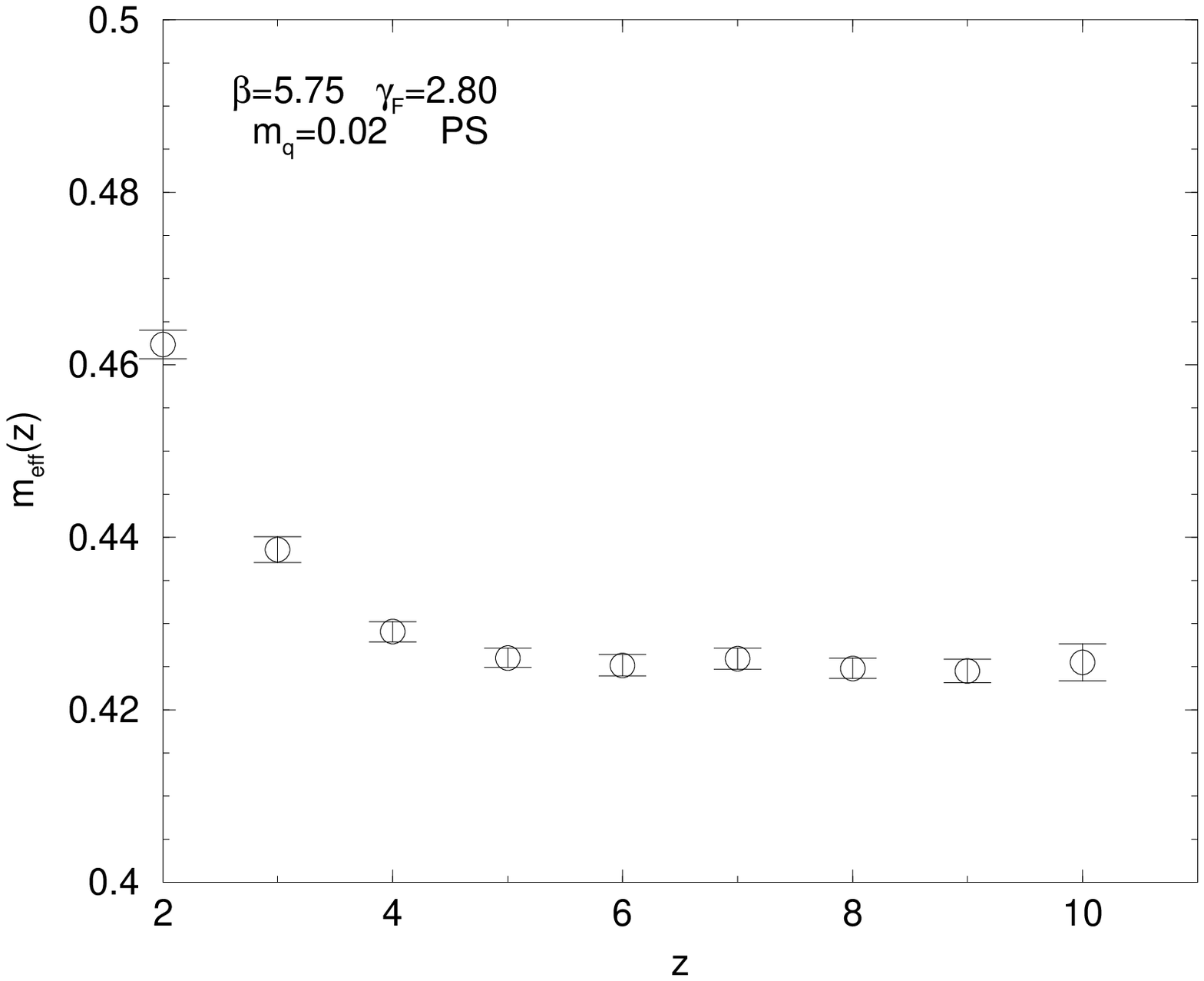}
\caption{
The effective mass plots for $m_q=0.02$ and $\gamma_F=2.8$
at $\beta=5.75$.
The left and right panels show the effective masses
in the $t$- (fine) and $z$- (coarse) directions, respectively.}
\label{fig:effective_mass}
\end{figure}

In the calibration, we use the pseudoscalar meson correlators
with the point source.
To perform the calibration along the two schemes described in
Sec.~\ref{sec:calibration_procedures},
we need the following two types of correlators for each set
of $m_q$ and $\gamma_F$:
\begin{itemize}

\item
 Correlators in the fine ($t$-) direction at finite momenta.
 We compute them at momenta $p_i=2\pi n_i /L_i$ ($i=x,y,z$)
 where $\vec{n}=$(0,0,0), (0,0,1), (0,0,2), (1,0,0), (1,0,1),
 (1,1,0), and (1,1,1).
 Note that $L_x = L_y = L_z/2$.

\item
 Correlators in the coarse ($z$-) direction at zero momentum.

\end{itemize}

To observe whether the correlator is dominated by a single
state and to determine a fit range,
we observe an effective mass $m_{eff}$ which is defined through
\begin{equation}
  \frac{M(t+2)}{M(t)}=
  \frac{\exp[-m_{eff}(t+2)]+\exp[-m_{eff}(N_t-t-2)]}
       {\exp[-m_{eff}t]+\exp(-m_{eff}(N_t-t)]} .
\end{equation}
The definition with lattice spacing $2a$ is adopted since
the correlators are composed of monotonously
and oscillatingly decaying modes.
An effective mass in $z$-direction is defined similarly.
Figure~\ref{fig:effective_mass} displays the effective masses
in the $t$- and $z$-directions at
$\beta=5.75$, $m_q=0.02$, and $\gamma_F=2.80$.

Figure~\ref{fig:effective_mass} shows that around $t=24$ and $z=6$
the correlators reach plateaus where
the contribution from the oscillating modes and excited
states is sufficiently reduced.
We choose a fit range for each quark mass, while the
same fit range is used for all the momenta, and fit the correlators
to a single exponential form.

Taking the ratio of the masses observed in the $t$- and
$z$-directions, the fermionic anisotropy $\xi_F^{(M)}$ is
determined for each input $\gamma_F$.

\begin{figure}[tb]
\includegraphics[width=7.0cm]{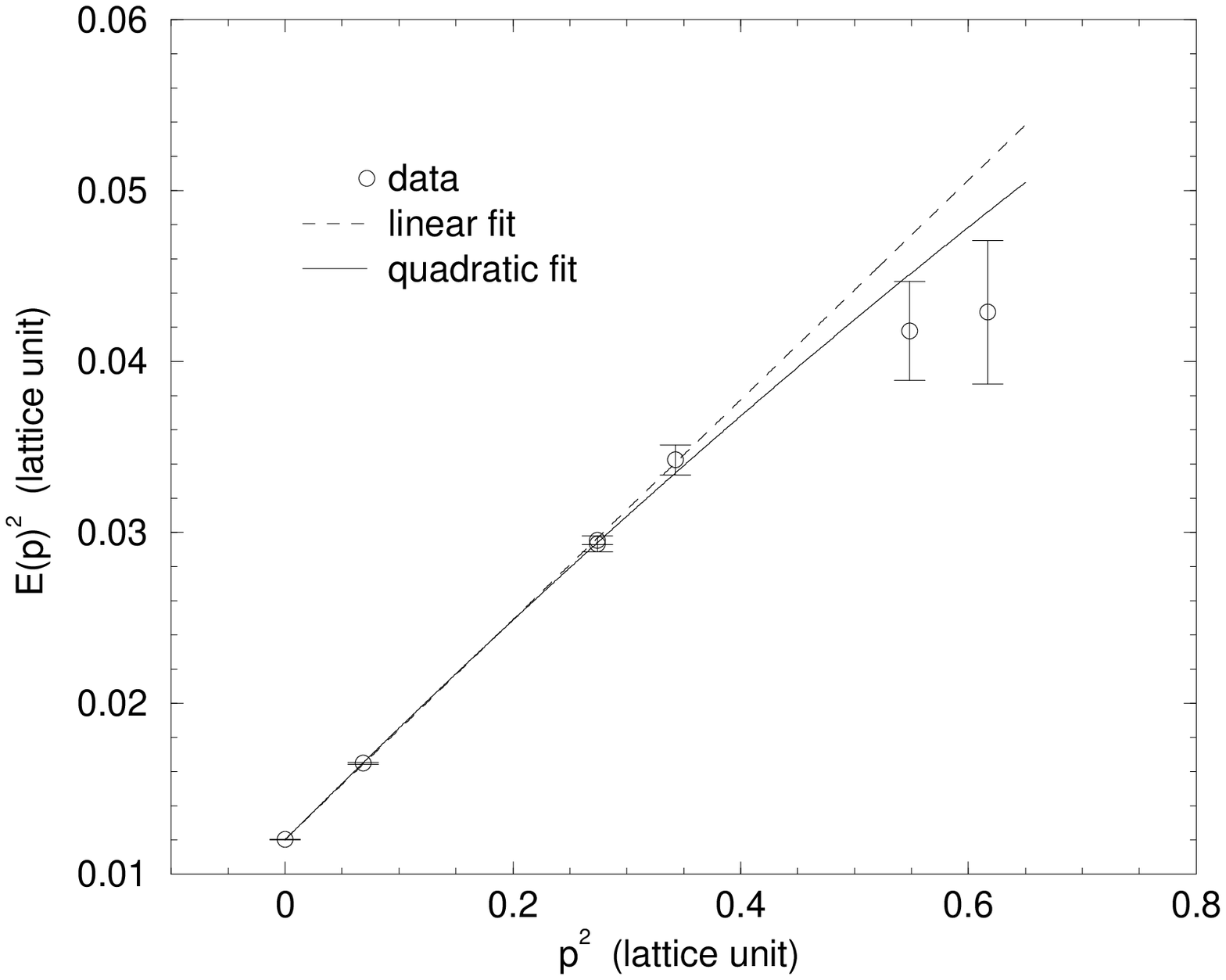}
\includegraphics[width=7.0cm]{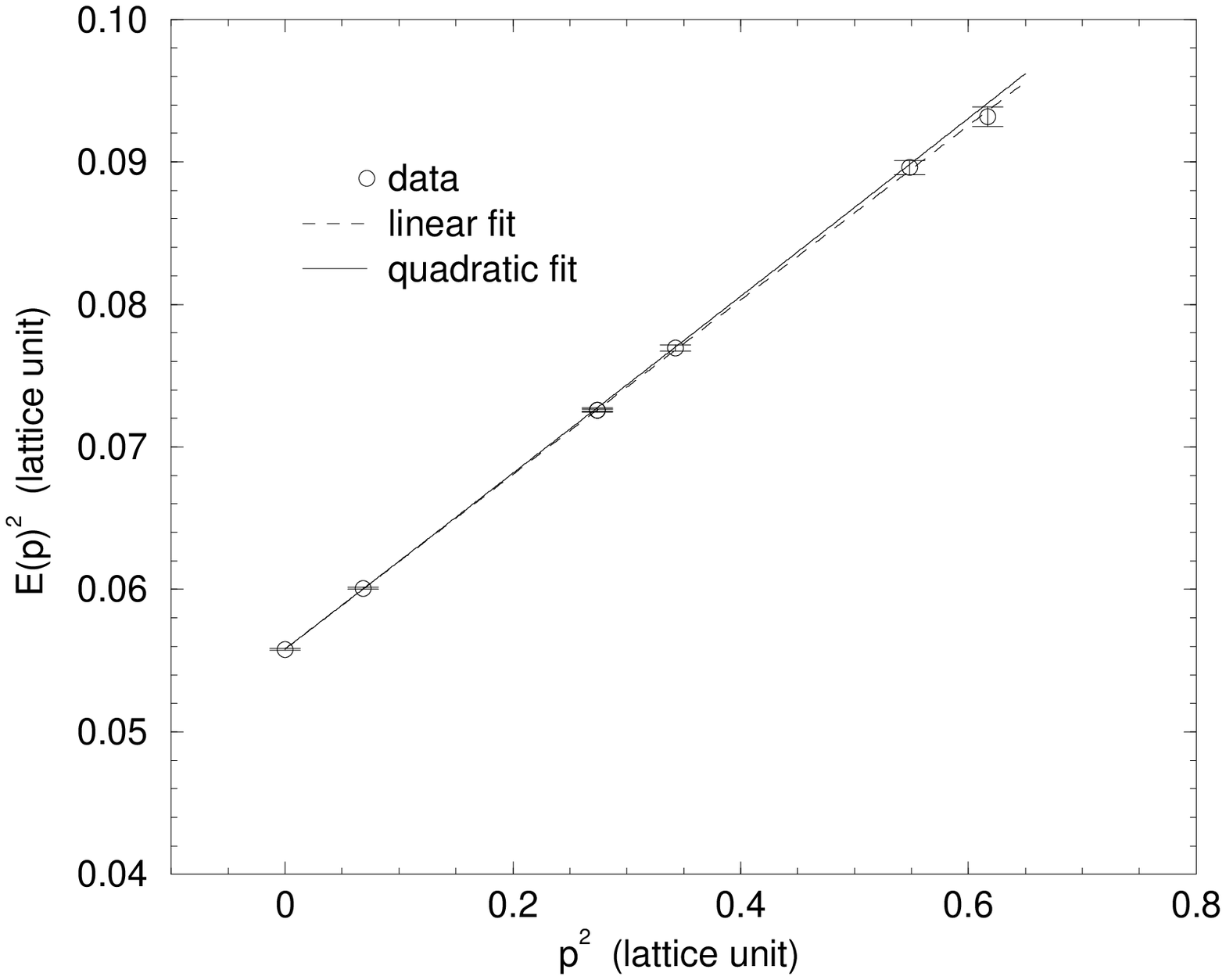}
\caption{
The dispersion relation at $\beta=5.75$.
The left and right panels display the results at
$(m_q,\gamma_F)=(0.02,2.8)$ and $(0.10,2.878)$, respectively.}
\label{fig:dispersion}
\end{figure}

Figure~\ref{fig:dispersion} shows the meson dispersion relation
for two quark masses at $\beta=5.75$.
$E(p)^2$ is almost a linear function of $p^2$ for small
momenta, and well fitted to a quadratic form in the whole measured
momentum region.
The same tendency is observed for all the quark masses and $\beta$'s
explored.
We apply a linear fit to the data of $E(p)^2$ with the smallest five
momenta, and determine the fermionic anisotropy $\xi_F^{(DR)}$
through its slope.
As displayed in Fig.~\ref{fig:dispersion}, the linear and quadratic
fits give almost the same results, while the former gives a smaller
statistical error.

\subsection{Determination of $\gamma_F^*$}

\begin{figure}[tb]
\includegraphics[width=7.0cm]{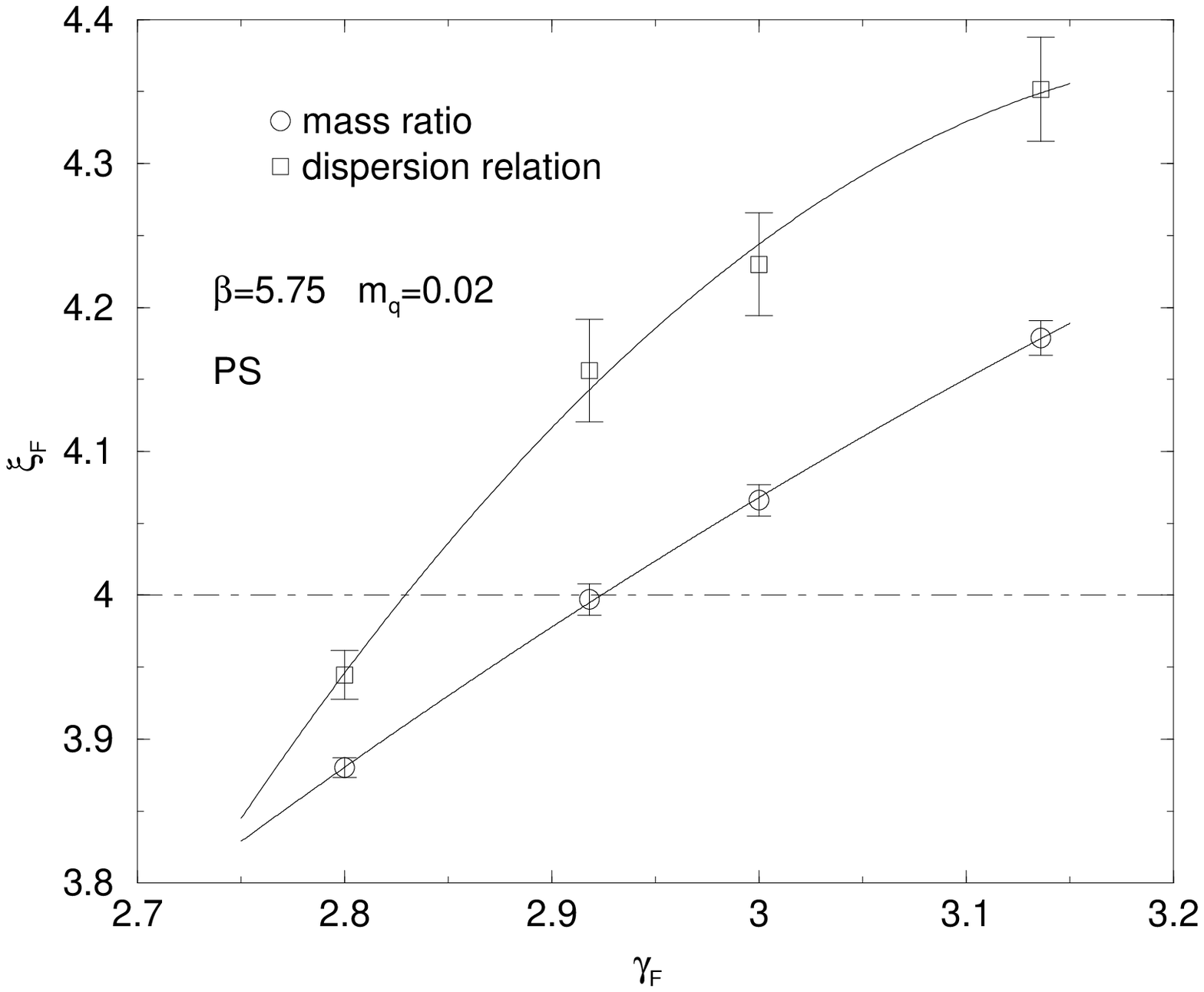}
\includegraphics[width=7.0cm]{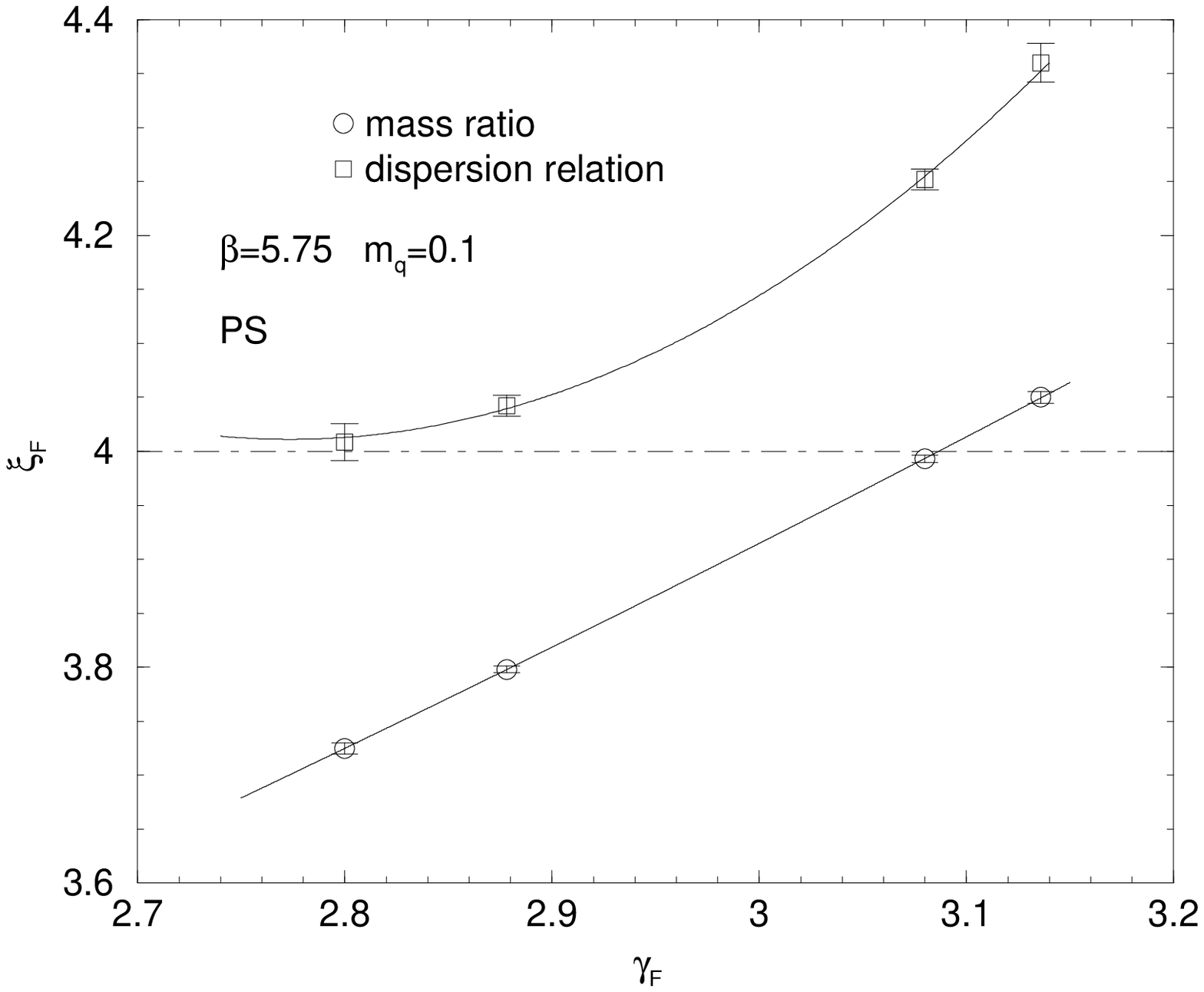}
\caption{
The dependences of $\xi_F$'s on $\gamma_F$ at $\beta=5.75$.
The left and right panels are for
$m_q=0.02$ and $0.10$, respectively.
The solid lines are the results of quadratic fits just for a guide
of eye, while in the determination of $\gamma_F^*$ the linear
fits using the data of $\xi_F$ near $\xi$ are used.}
\label{fig:gammaF_interpolation}
\end{figure}

\begin{table}[tb]
\caption{
The results of the calibration.
$\gamma_F^{*(M)}$ and $\gamma_F^{*(DR)}$ represent the tuned
anisotropy parameters in the mass ratio and dispersion relation
schemes, respectively.
The pseudoscalar quark masses at $\gamma_F^{*}$ are determined by
interpolation.}
\begin{center}
\begin{tabular}{cccccccc}
\hline \hline
$m_q$ & input $\gamma_F$ & 
         $\gamma_F^{*(M)}$  & $m_{PS}^{(t)}(\gamma_F^{*(M)})$  &
         $\gamma_F^{*(DR)}$ & $m_{PS}^{(t)}(\gamma_F^{*(DR)})$ \\
\hline
$\beta=5.75$ \\
0.50 & 3.578, 3.5761, 3.136, 2.80 &
        3.5822(21) & 0.46494(17) & 2.6777(47) & 0.59849(73) \\
0.40 & 3.471, 3.466, 3.136, 2.778 &
        3.4735(23) & 0.42055(17) & 2.7239(45) & 0.51765(61) \\
0.30 & 4.00, 3.353, 3.30, 2.80 &
        3.3590(27) & 0.36962(17) & 2.7700(47) & 0.43311(49) \\
0.20 & 3.219, 3.136, 2.80 &
        3.2308(35) & 0.30798(19) & 2.8073(60) & 0.34376(50) \\
0.10 & 3.136, 3.08, 2.878, 2.80 &
        3.0870(34) & 0.22486(16) & 2.821(11)  & 0.2398(10)  \\
0.05 & 3.136, 3.00, 2.987, 2.80 &
        2.9891(76) & 0.16339(18) & 2.8296(95) & 0.16979(37) \\
0.03 & 3.136, 3.00, 2.948, 2.80 &
        2.947(10)  & 0.12867(21) & 2.827(10)  & 0.13254(33) \\
0.02 & 3.136, 3.00, 2.918, 2.80 &
        2.930(13)  & 0.10630(22) & 2.831(10)  & 0.10845(29) \\
0.01 & 3.136, 3.00, 2.888, 2.80 &
        2.924(28)  & 0.07620(36) & 2.840(12)  & 0.07765(23) \\
\hline
$\beta=5.95$ \\
0.50 & 3.624, 3.578 &
       3.6246(16) & 0.45727(13) & 2.8513(94) & 0.54393(98)  \\
0.40 & 3.521, 3.471 &
       3.5198(19) & 0.41144(14) & 2.897(10)  & 0.47370(98)  \\
0.30 & 3.413, 3.353 &
       3.4113(25) & 0.35813(15) & 2.945(12)  & 0.39781(99)  \\
0.20 & 3.292, 3.219 &
       3.3012(36) & 0.29292(17) & 3.000(15)  & 0.31307(97)  \\
0.10 & 3.166, 3.080 &
       3.1958(41) & 0.20575(12) & 3.032(13)  & 0.21278(53)  \\
0.05 & 3.112, 2.987 &
       3.1460(91) & 0.14498(18) & 3.043(25)  & 0.14781(65)  \\
0.03 & 3.170, 3.100 &
       3.139(17)  & 0.11279(28) & 3.053(77)  & 0.1144(13)  \\
0.02 & 3.100, 2.918 &
       3.137(20)  & 0.09258(27) & 3.046(80)  & 0.0940(11)  \\
0.01 & 3.150, 3.100 &
       3.131(31)  & 0.06631(27) & 3.05(12)   & 0.0670(11)  \\
\hline
$\beta=6.10$ \\
0.30 & 3.21, 3.11 &
       3.4341(26) & 0.34436(17)  & 3.0174(95) & 0.38259(96)  \\
0.20 & 3.21, 3.11 &
       3.3309(35) & 0.27754(16)  & 3.068(10)  & 0.29493(74)  \\
0.10 & 3.21, 3.11 &
       3.2377(69) & 0.18959(19)  & 3.119(16)  & 0.19413(62)  \\
0.05 & 3.21, 3.16, 3.10 &
       3.199(16)  & 0.13037(24)  & 3.158(32)  & 0.13129(71)  \\
0.03 & 3.31, 3.21, 3.16 &
       3.199(29)  & 0.10009(25)  & 3.198(46)  & 0.10010(83)  \\
0.02 & 3.31, 3.21, 3.16 &
       3.210(45)  & 0.08155(29)  & 3.244(66)  & 0.08113(95)  \\
0.01 & 3.31, 3.21, 3.16 &
       3.250(80)  & 0.05798(34)  & 3.261(78)  & 0.05789(68)  \\
\hline\hline
\end{tabular}
\end{center}
\label{tab:calibration}
\end{table}

\begin{figure}[tb]
\includegraphics[width=7.0cm]{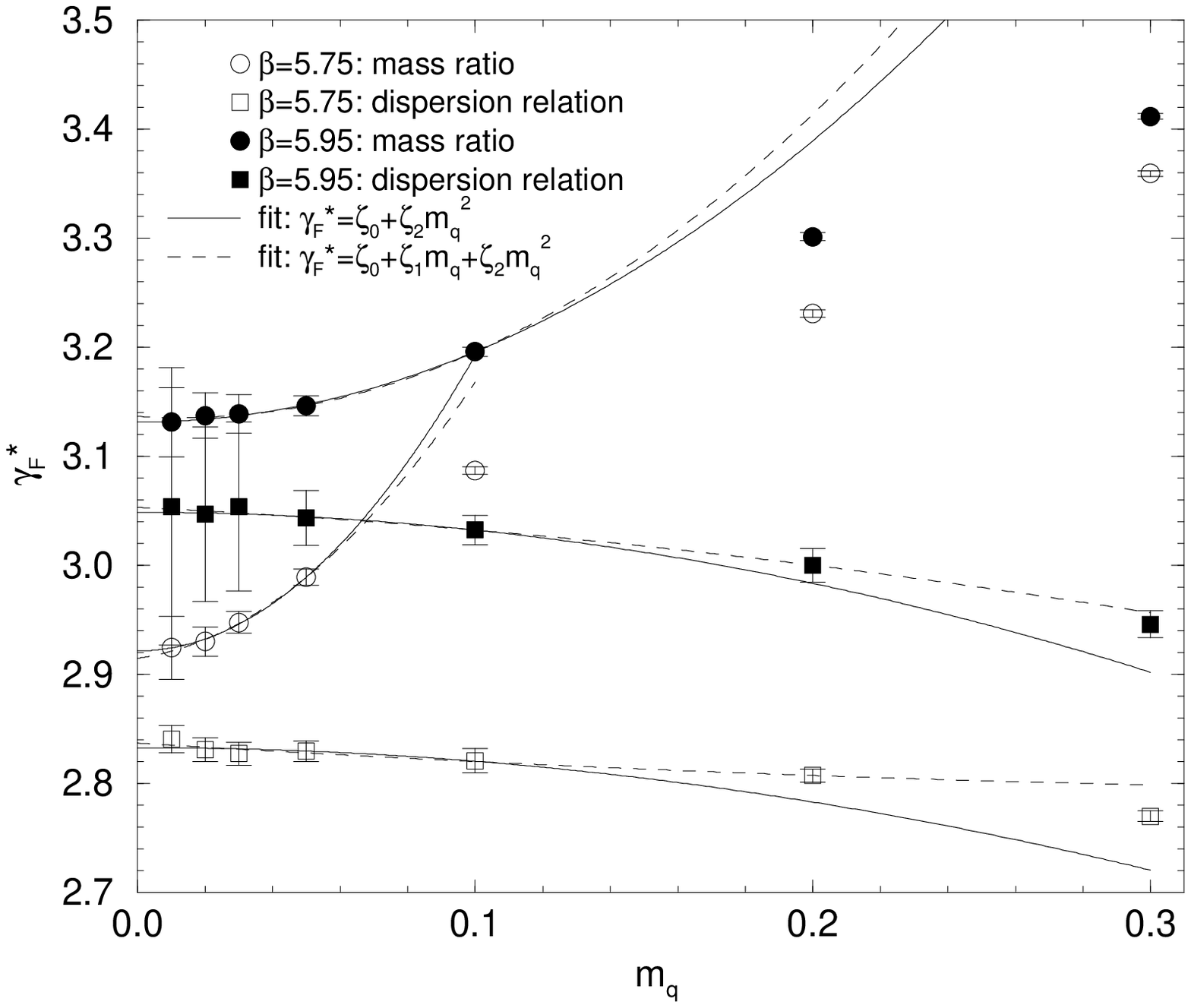}
\includegraphics[width=7.0cm]{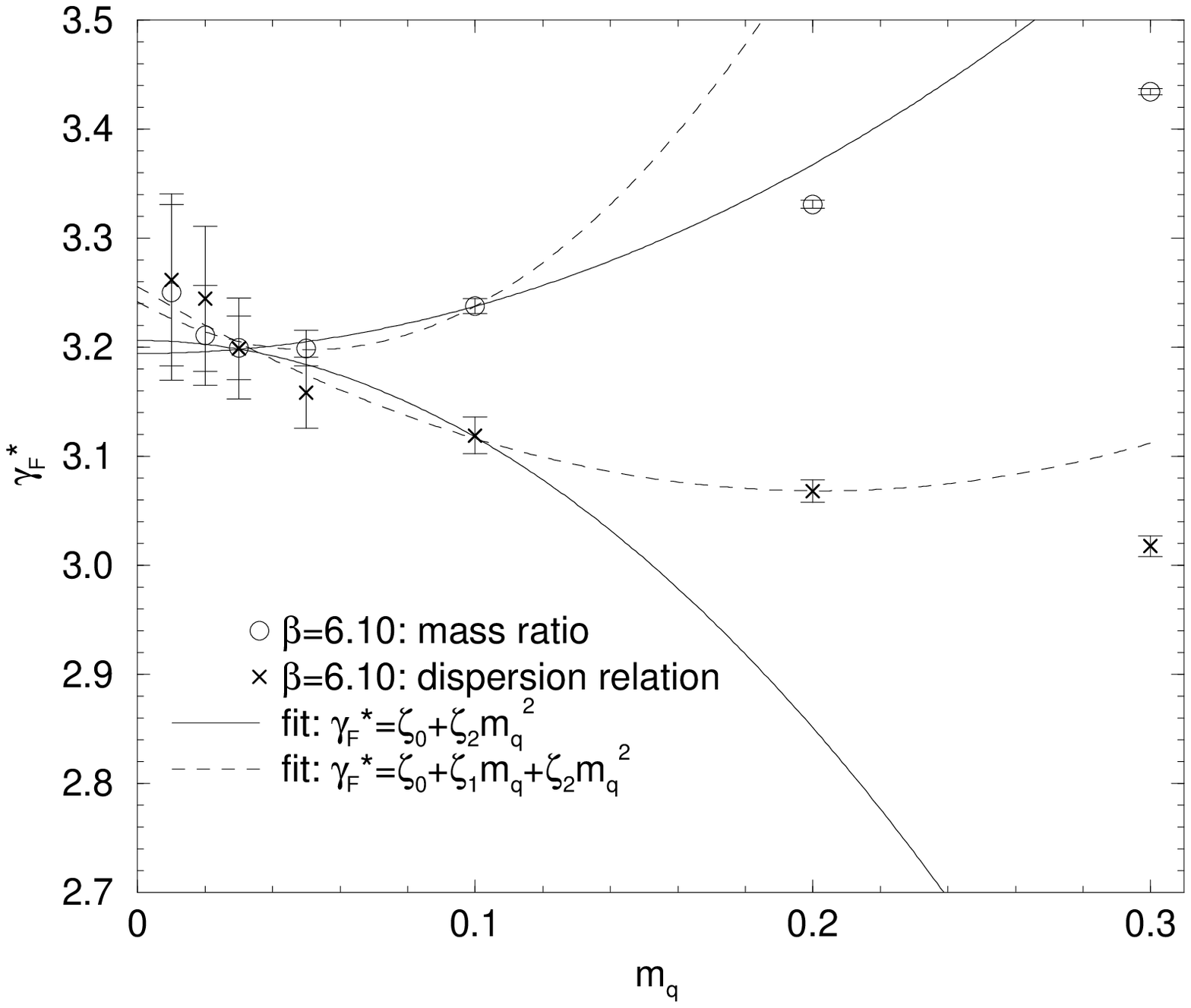}
\caption{
The results of the calibration.
The left panel displays the results at $\beta=5.75$ and
$\beta=5.95$.
The right panel shows the results at $\beta=6.10$.
The data are from Table~\ref{tab:calibration}.
The solid and dashed lines represent the results of fits which
are listed in Table~\ref{tab:calibration_fit}. }
\label{fig:calibration}
\end{figure}

We observe the fermionic anisotropies $\xi_F^{(M)}$ and
$\xi_F^{(DR)}$ for several values of $\gamma_F$ at each quark mass.
Figure~\ref{fig:gammaF_interpolation} displays typical $\gamma_F$
dependences of $\xi_F$'s at $\beta=5.75$.
This figure shows that the renormalized anisotropy is well represented
as a linear function of the bare anisotropy around $\xi_F = \xi$.
At $\beta=5.95$ and 6.10, we therefore in most cases measure the
correlators for two values of $\gamma_F$ at each quark mass.

Interpolating $\xi_F(\gamma_F)$ linearly in $\gamma_F$ to $\xi=4$, 
we can define a tuned anisotropy parameter $\gamma_F^*$.
To specify the tuning procedure, the $\gamma_F^*$ tuned
with the mass ratio (dispersion relation) scheme is denoted by
$\gamma_F^{*(M)}$ ($\gamma_F^{*(DR)}$).
The result of the interpolation is summarized in
Table~\ref{tab:calibration} and displayed in
Figure~\ref{fig:calibration}.

Figure~\ref{fig:calibration} shows large discrepancies between
the results of two calibration schemes in the large quark mass region.
The differences remain at the massless limit for $\beta=5.75$
and $5.95$, while disappears within errors at $\beta=6.10$.
This implies that the discrepancy is due to the finite
lattice spacing artifacts.
The analysis of the free quark propagator suggests that the quark
mass dependence of $\gamma_F^*$ on $m_q$ starts with a quadratic term
near the chiral limit in both the schemes.
This tendency is observed in the dispersion relation scheme
more clearly than the mass ratio scheme.
As the bare quark mass increases, 
$\gamma_F^{*(M)}$ increases while $\gamma_F^{*(DR)}$ decreases.
These quark mass dependences are consistent with the results
for the free quark case, Eqs.~(\ref{eq:calib_free_M}) and
(\ref{eq:calib_free_DR}).
For $\xi=4$, these expressions expect larger quark mass dependence
for $\gamma_F^{*(M)}$ than $\gamma_F^{*(DR)}$, which is also observed
in the numerical result.

The discrepancy between $\gamma_F^{*(M)}$ and $\gamma_F^{*(DR)}$ at
the chiral limit can be explained by the $O(a_\sigma^2)$ finite
lattice artifacts, such as
the flavor symmetry breaking effect, the $O((ap)^2)$ uncertainty
in the dispersion relation scheme, and so on.
The latter effect concerns the assumed form of the meson dispersion
relation as well as the fitting form for the energies at
finite momenta.
For the measurement of the mass in the $z$-direction,
the meson quantum number is composed of
the staggered quark fields on asymmetric cubes in a $z$-plane.
This leads to a different manifestation of the flavor symmetry breaking
effect from that of the mass in $t$-direction.
Although all these effects must disappear toward the continuum limit,
at finite lattice spacings they may cause intricate systematic effects.

\begin{table}[tb]
\caption{
The result of the fits of the tuned anisotropy parameter $\gamma_F^*$
to the form (\ref{eq:calibration_fit}).
In the case of linear fit, $\zeta_1$ is set to zero.
In the second column, `M' and `DR' represent the mass ratio
and dispersion relation schemes, respectively.
$N_{data}$ is the number of data points used in the fit.}
\begin{center}
\begin{tabular}{cccccccc}
\hline \hline
         &  \multicolumn{3}{c}{mass ratio} &
            \multicolumn{3}{c}{dispersion relation} \\
 $\beta$ & scheme & fit & $N_{data}$ &
           $\zeta_0$ &  $\zeta_1$ &  $\zeta_2$ & $\chi^2$ \\
\hline
5.75& M & linear   & 4 & 2.940(11)&    -      &   14.7(11)& 0.092 \\
    & M & quadratic& 4 & 2.915(42)&  0.4(19)  &    21.(23)& 0.001 \\
    & DR& linear   & 5 & 2.833(10)&    -      & $-$1.2(12)& 0.013 \\
    & DR& quadratic& 6 & 2.837(13)&$-$0.19(29)&    0.2(12)& 0.009 \\
\hline
5.95& M & linear   & 5 & 3.131(14)&    -      &    6.4(12)& 0.0007 \\  
    & M & quadratic& 5 & 3.137(40)& $-$0.2(12)&    7.9(78)& 0.0006 \\
    & DR& linear   & 5 & 3.049(38)&    -      & $-$1.6(35)& 0.0002 \\  
    & DR& quadratic& 6 & 3.053(71)& $-$0.2(11)& $-$0.6(38)& 0.0002 \\
\hline
6.10& M & linear   & 5 & 3.194(28)&    -      &    4.3(24)& 0.008 \\ 
    & M & quadratic& 5 & 3.242(81)& $-$1.7(21)&    17.(13)& 0.002 \\
    & DR& linear   & 5 & 3.206(45)&    -      & $-$8.9(40)& 0.021 \\  
    & DR& quadratic& 6 & 3.256(69)& $-$1.9(10)&    4.6(34)& 0.007 \\
\hline\hline
\end{tabular}
\end{center}
\label{tab:calibration_fit}
\end{table}

We determine $\gamma_F^*$'s at the vanishing quark mass by
fitting the data to a linear form in $m_q^2$ or quadratic form
in $m_q$,
\begin{equation}
  \gamma_F^* = \zeta_0 + \zeta_1 m_q + \zeta_2 m_q^2,
\label{eq:calibration_fit}
\end{equation}
where $\zeta_1$ is set to zero in the linear fit.
The numbers of data used in the fits are selected appropriately.
The result of the fit is listed in Table~\ref{tab:calibration_fit},
and displayed in Fig.~\ref{fig:calibration}.
For both the schemes, the linear form in $m_q^2$ seems to represent
well the data in the light quark mass region.
Although the quadratic form does not seem to work properly,
the $\gamma_F^*$ at the chiral limit is close to the result of
the linear fit.

\subsection{Pion mass and the chiral limit}

\begin{figure}[tb]
\includegraphics[width=7.0cm]{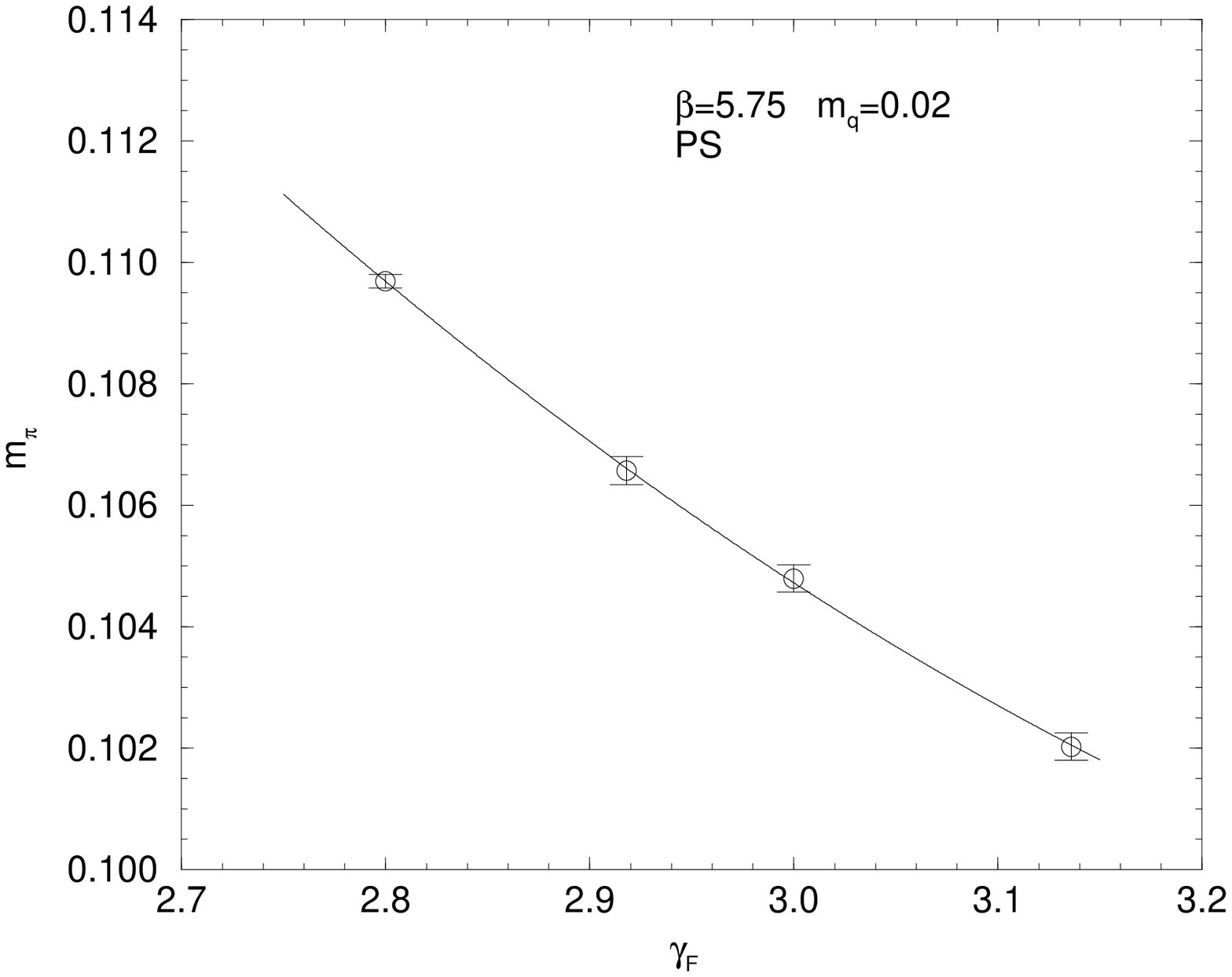}
\includegraphics[width=7.0cm]{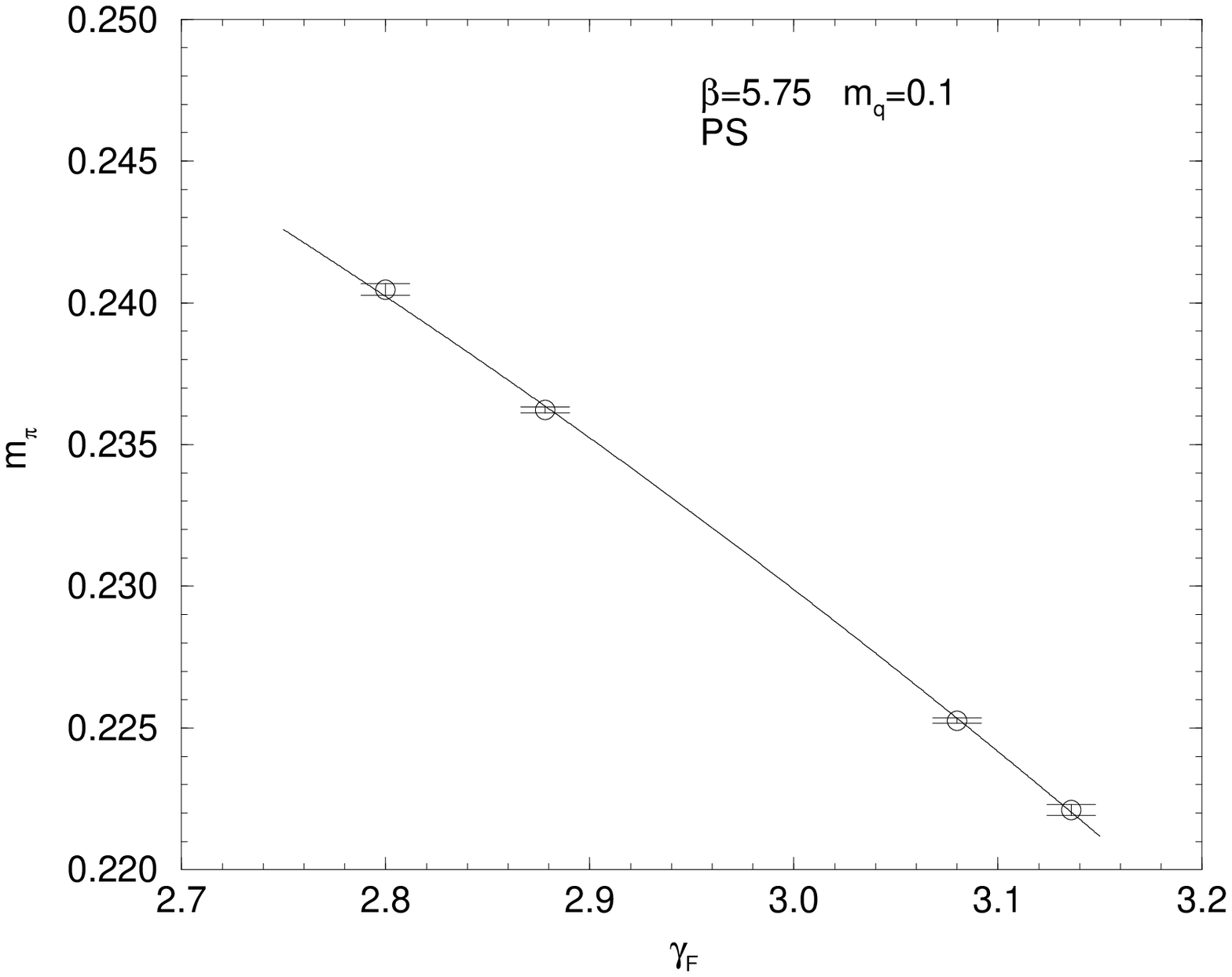}
\caption{
The dependence of $m_{PS}$ on $\gamma_F$ at $\beta=5.75$.
The left and right panels show the results at
$m_q=0.02$ and $0.10$, respectively.
The solid lines represent the results of quadratic fits.}
\label{fig:mpi_interpolation}
\end{figure}

\begin{figure}[tb]
\centerline{
\includegraphics[width=9.0cm]{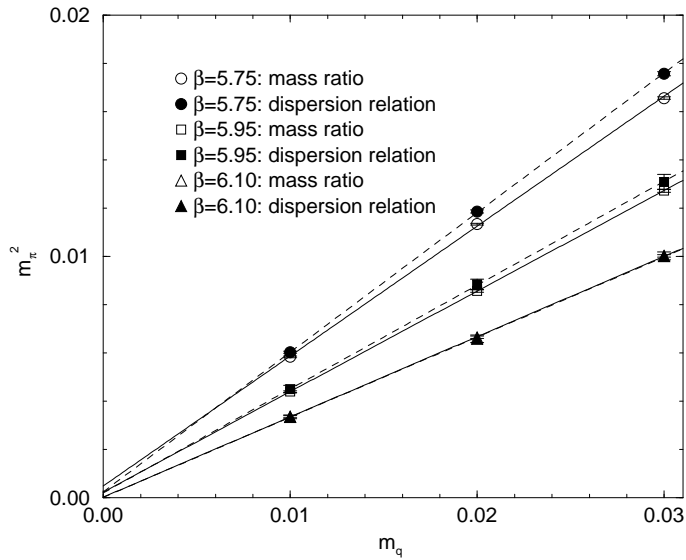}}
\caption{
The dependence of $m_{PS}^2$ on the bare quark mass $m_q$.
The solid and dashed lines represent the results of linear fits.}
\label{fig:mpi_gammaF}
\end{figure}

Here we verify that the pseudoscalar meson mass satisfies
the PCAC relation with the calibrated anisotropy parameter.
Firstly we need to interpolate the pseudoscalar meson mass
to $\gamma_F=\gamma_F^*$.
Figure~\ref{fig:mpi_gammaF} shows typical $\gamma_F$ dependences of
PS meson masses at $\beta=5.75$.
The PS meson mass is well represented by a linear form
in the vicinity of $\gamma_F^*$.
The result of the linear interpolation is listed
in Table~\ref{tab:calibration}.

Figure~\ref{fig:mpi_gammaF} displays the $m_q$ dependence of
the pseudoscalar meson mass squared.
The proportionality of $m_{PS}^2$ to $m_q$ well holds for both
the calibration schemes.
This is also verified by linear fits whose results are listed in
Table~\ref{tab:PCAC}.
In either scheme, the intercept is quite small and becomes
consistent with zero within the error as $\beta$ increases.
The difference between the two schemes also decreases toward
the continuum limit.

\begin{table}[tb]
\caption{
The results of linear fits of $m_{PS}^2$ in $m_q$ at the tuned
anisotropy $\gamma_F^*$.}
\begin{center}
\begin{tabular}{ccccccc}
\hline \hline
         &  \multicolumn{3}{c}{mass ratio} &
            \multicolumn{3}{c}{dispersion relation} \\
 $\beta$ & $m_\pi^2(0)$ & $dm_\pi^2/dm_q$ & $\chi^2$ &
           $m_\pi^2(0)$ & $dm_\pi^2/dm_q$ & $\chi^2$  \\
\hline
 5.75 & 0.000493(38) & 0.5382(17) & 8.14 &
        0.000251(49) & 0.5784(36) & 0.535 \\
 5.95 & 0.000236(36) & 0.4164(22) & 0.0407 &
        0.00021(20)  & 0.430(13)  & 0.0214 \\
 6.10 & 0.000028(43) & 0.3325(16) & 0.458 &
        0.000025(91) & 0.3317(70) & 0.319 \\
\hline\hline
\end{tabular}
\end{center}
\label{tab:PCAC}
\end{table}

\subsection{Summary of calibration}

In the next section, we investigate the meson spectrum of
the anisotropic staggered quark.
For this purpose, it is convenient to select one of the results
of calibrations as our main result for $\gamma_F^*$ and
compute the spectrum with this value.
The differences with the other choices are treated as a systematic
uncertainty whose effect on the spectrum should be investigated.

As such a representative, we adopt the $\gamma_F^*$ in the massless
limit determined by the linear fit in $m_q^2$ in the dispersion
relation scheme for the following reasons:
As apparent in Fig.~\ref{fig:calibration}, $\gamma_F^{*(DR)}$
less depends on the quark mass than $\gamma_F^{*(M)}$.
In the light quark mass region we can use the value of
$\gamma_F^*$ at the chiral limit instead of the direct results on
those quark masses.
The statistical error of the former can be reduced less than those
of the latter.

Estimate of the statistical error of $\gamma_F^*$ at massless limit
is provided by the statistical error of the fit result for $\zeta_0$.
On the other hand, the discrepancy between the two calibration
schemes gives a typical size of the systematic uncertainty.

To summarize, we select the following values as a representative
result of the calibration:
\begin{eqnarray}
 \beta=5.75:&\ \ & \gamma_F^* = 2.83(1)(11) , \nonumber \\
 \beta=5.95:&\ \ & \gamma_F^* = 3.05(4)(11) ,
\label{eq:calib_result} \\
 \beta=6.10:&\ \ & \gamma_F^* = 3.21(5)(5) , \nonumber
\end{eqnarray}
where the first and second parentheses denote the statistical
and systematic errors, respectively.
The latter does not include the quark mass dependence of $\gamma_F^*$.
At $\beta=6.10$, the domination of the statistical error around
the chiral limit disable us from estimating the systematic error
in the same way as other $\beta$'s.
From the conservative point of view, we substitute the statistical
error for the systematic one, since the latter is at most of the
same size as the former.

Finally let us compare the result of calibration with the
mean-field estimate \cite{LM93}.
The mean-field improvement is performed by replacing the
link variables as $U_i\rightarrow U_i/u_\sigma$ ($i=1,2,3$)
and $U_4 \rightarrow U_4/u_\tau$.
The mean-field estimate of $\gamma_F^*$ in the chiral
limit results in
\begin{equation}
 \gamma_F^{*(MF)} = \xi \cdot (u_\sigma/u_\tau).
\end{equation}
From the values quoted in Table~\ref{tab:parameters}, 
$\gamma_F^{*(MF)}$ is determined as
3.0878(8)\footnote{At $\beta=5.75$ the value has additional
   systematic error since the mean-field values are obtained
   at slightly different $\gamma_F$.}
($\beta=5.75$), 3.2017(4) ($\beta=5.95$),
and 3.2558(4) ($\beta=6.10$).
The result of numerical simulation approaches to the mean-field
estimate from below as $\beta$ increases.
This suggests that the mean-field value of $\gamma_F^*$
provides a good guide in the calibration.
This feature is helpful in the calibration of anisotropic lattices
with dynamical quarks.


\section{Spectroscopy on anisotropic lattices}
 \label{sec:spectroscopy}

In this section, we compute the meson spectrum using the
anisotropic staggered quark action tuned in the previous section.
As noted at the end of Sec.~\ref{sec:calibration},
we use the values of $\gamma_F^*$ quoted in
Eq.(\ref{eq:calib_result}) in the simulation and
investigate how the uncertainties in $\gamma_F^*$ affect
the spectrum.
In practice, we compute the masses at two $\gamma_F$'s and
linearly interpolate them to $\gamma_F^*$
in Eq.(\ref{eq:calib_result}).
Simultaneously the response of the mass to the change
of $\gamma_F$ is obtained.

The numerical simulation is performed at the same three $\beta$'s
as in the calibration.
At $\beta=5.75$, the lattice size is the same as in the calibration.
At $\beta=5.95$ and 6.10, we use the lattices with half size
in the $z$-direction, which were used in Ref.~\cite{Aniso01b}.
The numbers of configurations are 224, 400, and 200 for $\beta=5.75$,
5.95, and 6.10, respectively.

We measure the meson correlators with the wall source \cite{GGKS91}.
The correlators measured were described in Sec.~\ref{sec:action}.
The gauge configurations are fixed to the Coulomb gauge.

The complication in analyzing the spectrum of the staggered mesons
arises from the oscillating modes contained in the correlators.
This makes fit more involved than other quark formulations.
In  extracting the meson masses, we apply the constrained curve
fitting \cite{Lepage02} to the fit of the correlator to
the form (\ref{eq:fit_form}).
In principle the constrained curve fitting enables a fit
of correlators to multipole forms with arbitrary number of terms.
In practice, however, we find that the mutipole fitting does not
produce a stable result partially because of the large fluctuations.
We therefore fit the data to the form (\ref{eq:fit_form})
in relatively narrow fit range where excited modes are
sufficiently reduced according to the observation of
the effective mass plot.

\begin{table}[tb]
\caption{
The meson spectra from the correlators with the wall source.
The meson masses at $m_q=0$ are determined by linear fits
using the data at lightest three $m_q$'s.}
\begin{center}
\begin{tabular}{ccccccc}
\hline \hline
 $m_q$ &  $\pi$ &  $\tilde{\pi}$ & $\pi_3$ & $\tilde{\pi}_3$ &
 $\rho_6^A$ & $\rho_6^B$  \\
\hline
$\beta=5.75$ \\
0.10 & 0.23846(10) &  0.24874(12) &  0.29410(40) &  0.29494(48) &
       0.34139(83) &  0.3413(13) \\
0.05 & 0.16935(12) &  0.17916(32) &  0.22577(43) &  0.22749(69) &
       0.28222(94) &  0.2831(16) \\
0.03 & 0.13211(14) &  0.14245(43) &  0.19110(39) &  0.1941(11) &
       0.25546(96) &  0.2557(16) \\
0.02 & 0.10854(15) &  0.11997(49) &  0.17199(48) &  0.1753(16) &
       0.2408(13)  &  0.2406(23) \\
0.01 & 0.07755(15) & 0.09159(60)  &  0.14948(68) &  0.1569(22) &
       0.2209(22)  &  0.2207(34) \\
0.   & 0.01726(67) & 0.0495(15)   &  0.12387(98) &  0.1336(34) &
       0.2064(28)  &  0.2054(42) \\
\hline
$\beta=5.95$ \\
0.10 & 0.21169(~8) &  0.21571(12) &  0.23084(17) &  0.23136(24) &
       0.25809(47) &  0.25816(63) \\
0.05 & 0.14754(10) &  0.15056(14) &  0.16469(20) &  0.16529(37) &
       0.20203(53) &  0.20163(73) \\
0.03 & 0.11429(13) &  0.11704(17) &  0.13182(24) &  0.13222(50) &
       0.17759(74) &  0.1767(11) \\
0.02 & 0.09380(15) &  0.09660(23) &  0.11250(28) &  0.11296(61) &
       0.16390(94) &  0.1633(17) \\
0.01 & 0.06712(13) &  0.07062(36) &  0.08950(48) &  0.0905(14)  &
       0.1468(14)  &  0.1476(24) \\
0.   & 0.01503(60) &  0.0250(13)  &  0.05746(87) &  0.0589(27)  &
       0.1330(18)  &  0.1341(30) \\
\hline
$\beta=6.10$ \\
0.10 & 0.19007(~9) &  0.19200(10) &  0.19945(14) &  0.19986(21) &
       0.21820(28) &  0.21745(27) \\
0.05 & 0.13006(10) &  0.13114(19) &  0.13759(25) &  0.13820(29) &
       0.16415(33) &  0.16414(34) \\
0.03 & 0.09990(10) &  0.10070(21) &  0.10733(20) &  0.10767(34) &
       0.14073(39) &  0.14089(53) \\
0.02 & 0.0816(11)  &  0.08239(22) &  0.08985(21) &  0.08957(41) &
       0.12870(47) &  0.12837(75) \\
0.01 & 0.0584(13)  &  0.05924(32) &  0.06889(20) &  0.06910(57) &
       0.11589(72) &  0.1134(13) \\
0.   & 0.0112(10)  &  0.0135(19)  &  0.03684(47) &  0.0360(13)  &
       0.10381(87) &  0.1012(16) \\
\hline\hline
\end{tabular}
\end{center}
\label{tab:spectrum}
\end{table}

\begin{figure}[tb]
\centerline{
\includegraphics[width=7.0cm]{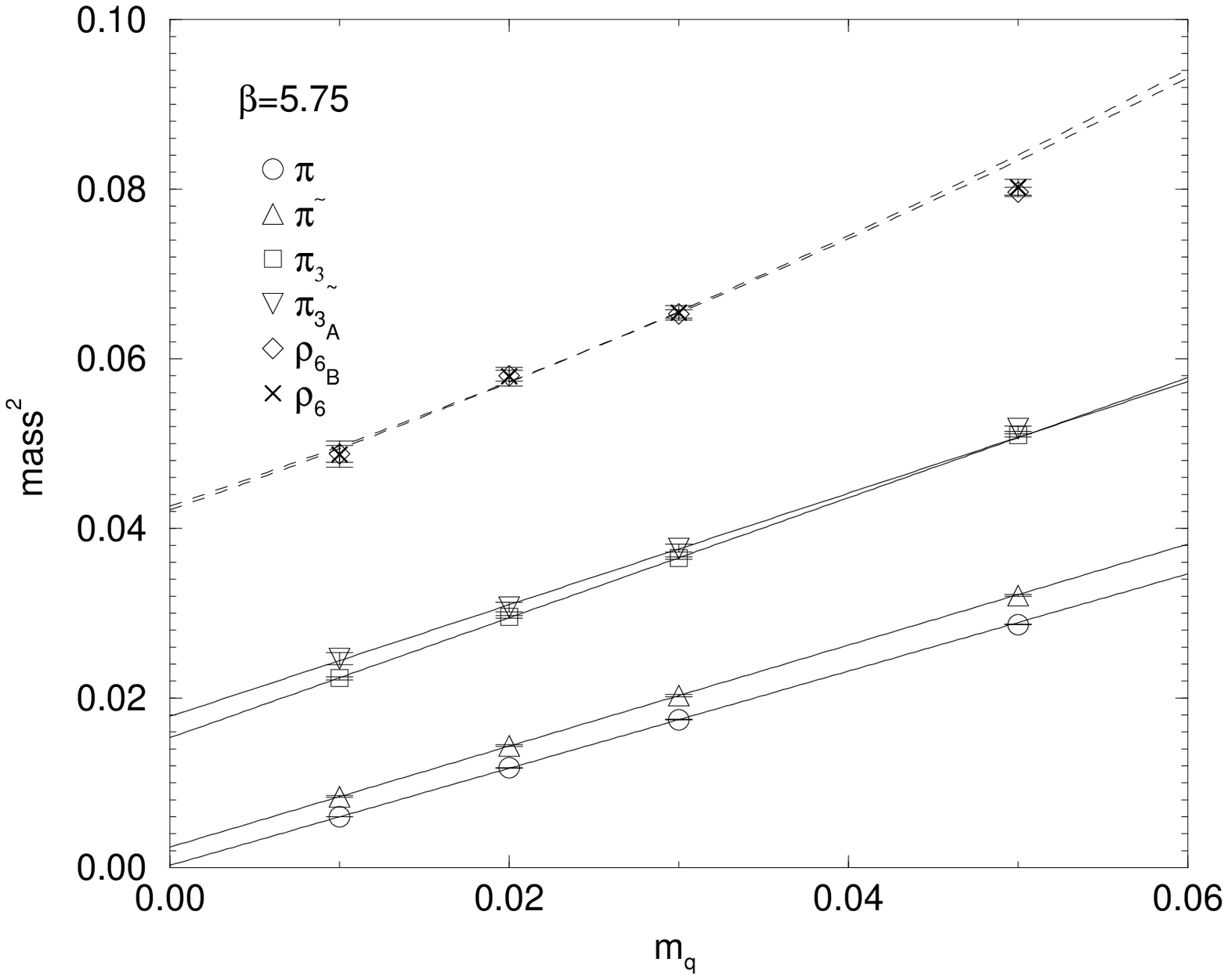}
\includegraphics[width=7.0cm]{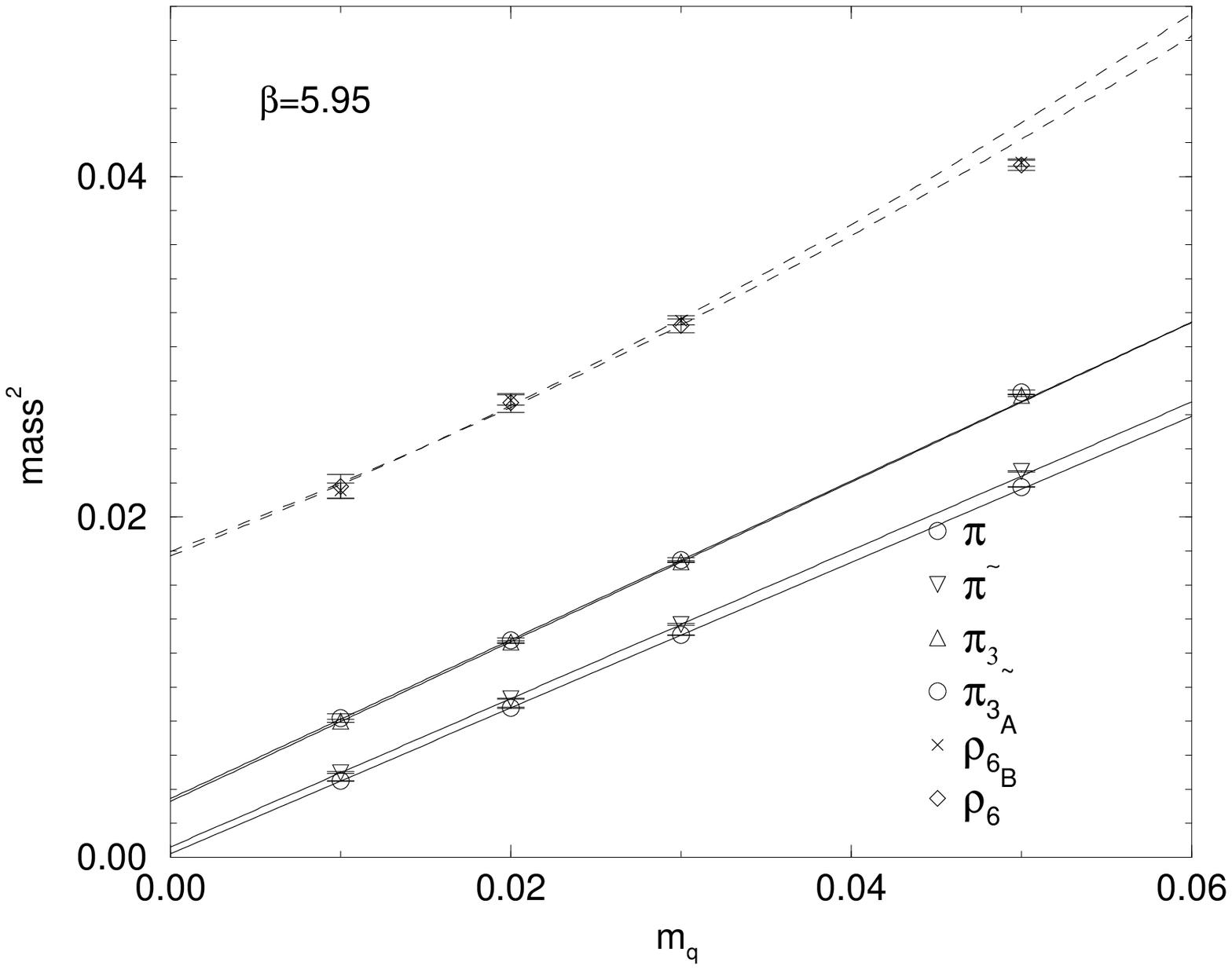}}
\centerline{
\includegraphics[width=7.0cm]{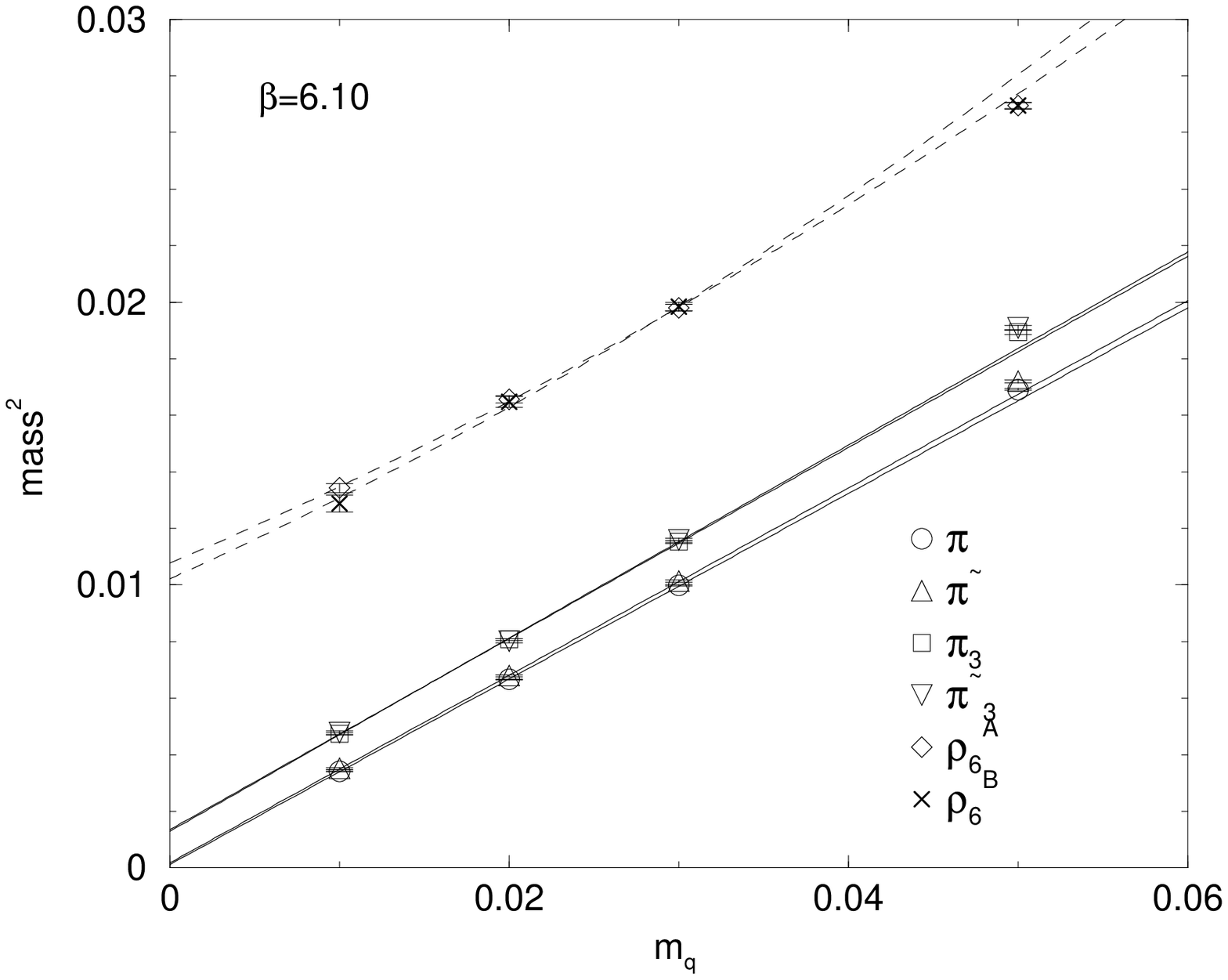}
\includegraphics[width=7.0cm]{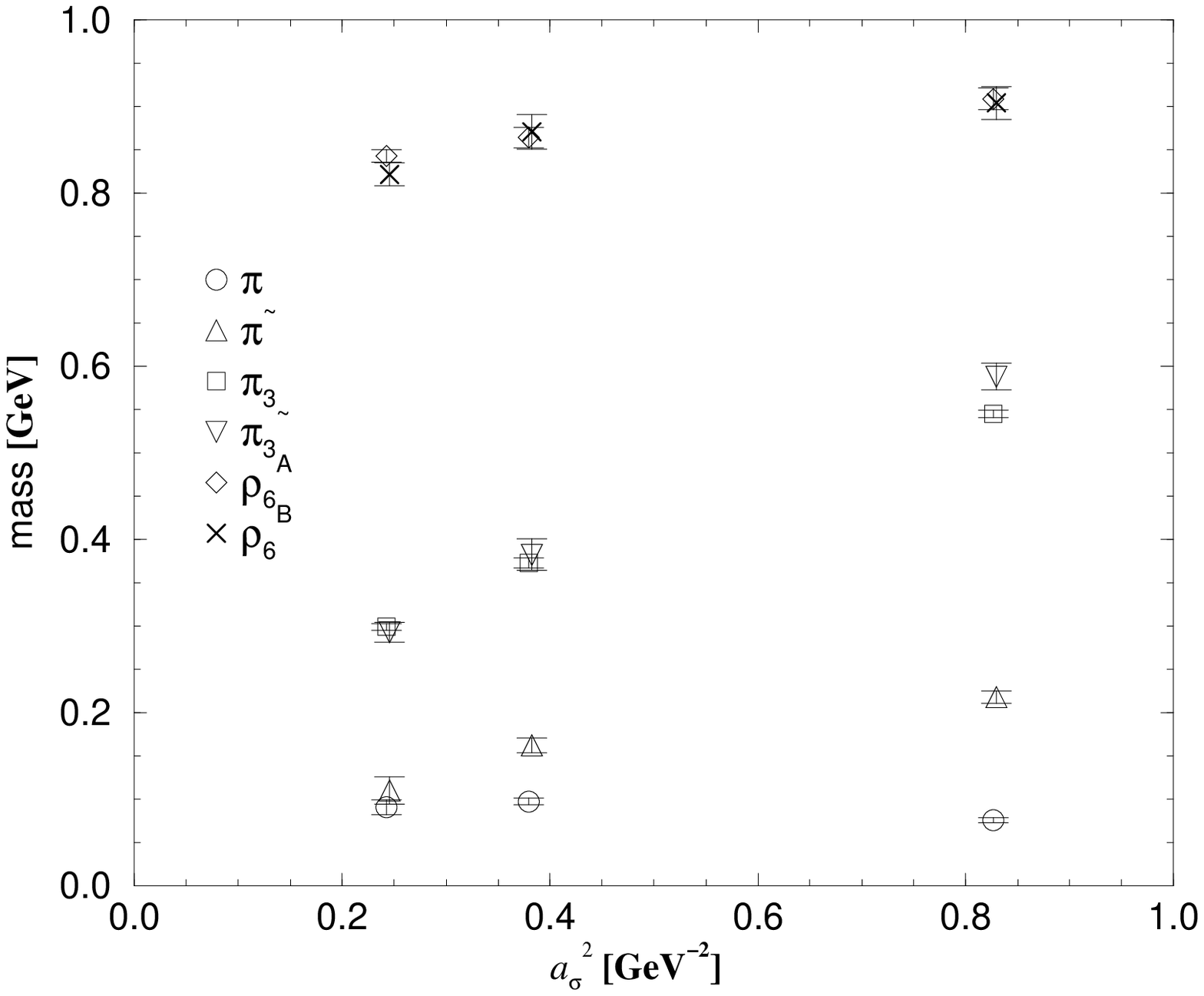}}
\caption{
Meson spectra at $\beta=5.75$ (top-left),
5.95 (top-right), and 6.10 (bottom-left). 
The curved lines represent the results of linear fits
using the data at the lightest three quark masses.
The bottom-right panel shows the meson masses in the chiral limit
in physical units.}
\label{fig:spectrum}
\end{figure}

The result for the spectrum is listed in Table~\ref{tab:spectrum}.
The quoted errors are statistical errors evaluated with
jackknife method.
The meson masses for the four lightest quark masses are displayed
in Figure~\ref{fig:spectrum}.
The figure shows that $\pi$ and $\tilde{\pi}$,
$\pi_3$ and $\tilde{\pi}_3$, $\rho_6^A$ and $\rho_6^B$ are
respectively degenerate.
All pionic channels approach to the masses of $\pi$, the Goldstone
pion channel, as $\beta$ increases.
This behavior is a signal of flavor symmetry restoration toward
the continuum limit.
The size of flavor symmetry breaking effect is estimated with
the mass of, say, $\pi_3$.
Our result of the $\pi_3$ mass is the same size as the result
of Ref.~\cite{GGKS91} at $\beta=6.0$ ($a\simeq 2$ GeV)
on an isotropic lattice.

To determine the masses at the chiral limit, we fit the
lightest three masses in each channel to the following forms:
\begin{eqnarray}
 m_\pi^2(m_q) &=& m_\pi^2(0) + b_\pi m_q
  \hspace{0.5cm}
  \mbox{($\pi$, $\tilde{\pi}$, $\pi_3$, $\tilde{\pi}_3$),} \\
 m_\rho(m_q) &=& m_\rho(0) + b_\rho m_q
  \hspace{0.57cm}
  \mbox{($\rho_6^A$, $\rho_6^B$).}
\end{eqnarray}
The results of the fits are also listed in Table~\ref{tab:spectrum}
and shown in Figure~\ref{fig:spectrum}.
The above forms well represent the data.
We also perform a fit of meson mass squared to a quadratic
form in $m_q$ using the four smallest meson masses in each
channel.
The results of the latter fits are consistent with those with
the former linear fits.

In the Goldstone pion channel, a small mass remains even in
the chiral limit.
This is considered due to the finite volume effect and
the uncertainty in $\gamma_F^*$.
The masses of other pionic channels in the chiral limit 
decrease as $\beta$ increases.
This behavior is shown more apparently in the bottom-right 
panel in Figure~\ref{fig:spectrum}, which displays the meson masses
in the chiral limit in physical units as the function of $a_\sigma^2$.
Since the staggered action contains $O(a_\sigma^2)$ systematic
uncertainty, the flavor symmetry breaking should disappear
linearly in $a_\sigma^2$.
Our result is consistent with this prediction, while the finite
lattice artifact is sizable even at our largest $\beta$.
Rather large lattice artifact is also expected from the
discrepancy between two calibration schemes examined in this work.
The $\rho$ meson channel is also consistent with the
expected behavior.
The masses of $\rho$ in the chiral limit seems to approach
the experimental value, $m_\rho=770$ MeV, within the
$O(10\%)$ error of the quenched approximation.

To investigate how the uncertainties of $\gamma_F$ affect
the masses, we evaluate
\begin{equation}
  R_H = \frac{(dm_H/d\gamma_F)}{m_H}
\end{equation}
assuming linear dependence of $m_H$ in $\gamma_F$
in the vicinity of $\gamma_F^*$.
The subscript $H$ specifies the channel.
As a general tendency, $R$ increases as quark mass decreases.
This means that the accurate determination of $\gamma_F^*$
becomes increasingly important as quark mass decreases.
$R$ is in general negative.
In pionic channels the absolute values of $R$ are 1--2
at $\beta=5.75$ and 5.95 and around 1 at $\beta=6.10$
for $m_q=0.01$, while at $m_q=0.1$ it decreases less than half
the values at $m_q=0.01$.
By adding the two errors in Eq.~(\ref{eq:calib_result})
in quadrature, the error of $\gamma_F^*$ is evaluated as
0.11, 0.12, and 0.07 for $\beta=5.75$, 5.95, and 6.10,
respectively.
At $m_q=0.01$ therefore the uncertainty in the meson mass
due to the uncertainty in $\gamma_F^*$ amounts to 20\% at
$\beta=5.75$ and 5.95, and 7\% at $\beta=6.10$.
These uncertainties explain the nonvanishing masses of
Goldstone pion at the chiral limit.
For the vector channels, the values of $R$ are about 1.5 times
larger than the pionic channels.
These quite significant effects call for precise
calibrations in order to use the anisotropic staggered quark
action in practical simulations.


\section{Conclusion}
 \label{sec:conclusion}

In this paper, we performed a calibration of the staggered quark
action on quenched anisotropic lattices with the renormalized
anisotropy $\xi=4$.
As the calibration procedures, we adopted two schemes which define
the fermionic anisotropy through the masses in the fine and coarse
directions, and through the meson dispersion relation.
At values of $\beta$ explored in this work, these two schemes
produce inconsistent result even in the chiral limit except
for our highest $\beta$ ($a_\sigma\simeq 2$ GeV).
Although the discrepancy seems to disappear toward the continuum
limit, these rather large discrepancy should be regarded as a
source of significant systematic uncertainties.

In the second part of this paper, we performed a meson
spectroscopy with the wall source using the values of
$\gamma_F^*$ determined in the calibration.
The result is consistent with the result on an isotropic lattice
\cite{GGKS91}.
Toward the continuum limit, the flavor symmetry breaking effect
seems to disappear.
However, the uncertainty of $\gamma_F$ causes severe effect
on the meson masses when the quark mass approaches the
chiral limit.
For precision studies using staggered anisotropic quarks,
precise calibration is indispensable.

This work displayed how the staggered quark action is realized
on anisotropic lattices in the quenched approximation.
Except for the rather large artifact signaled by
the discrepancy between the two calibration schemes, no unreasonable
result was found.
However, there are numbers of subject which should be
examined still in the quenched simulations.
It is important to study in more detail how the
discrepancy between the calibration schemes disappear as one
approaches the continuum limit and to quantify the effect on
the observables such as the hadron masses and decay constants.
The other important subject is to examine how the various
channels behave when they are measured in the coarse direction
or boosted into a finite momentum state.
These studies are important to investigate the structure
of the staggered quark formulation as well as for applications
to the calculations for which the fine temporal cutoff is
crucial.
For intermediate range of lattice cutoffs,
development of improved staggered quark action is also
important to avoid the large lattice artifact.
The results of this paper provide fundamental information
for such further studies.


\section{Toward dynamical simulations}
 \label{sec:dynamical}

In this final section, we discuss what we learn from the quenched
results in this work toward the dynamical simulations.
Since in dynamical simulations the generation of gauge configurations
requires resources, one wants to finish the calibration as quickly
as possible.
In this sense, the mass ratio scheme is economical since the
statistical error in $\xi_F$ is smaller than the dispersion relation
scheme.
However, larger quark mass dependence of $\gamma_F$ is inconvenient
to survey the global dependence of $\gamma_F^*$ on $\beta$ and
$m_q$.
In this sense, the dispersion relation scheme has an advantage.
There is another advantage in the latter: one needs to have
a large extent only in the temporal (fine) direction.
The spatial direction can be kept modest size with which the
momentum modes are not very large.
One can also use the extended source field to reduce the
statistical fluctuations.
For these reasons, we realize that the dispersion relation scheme is
more suitable for the calibration of dynamical lattices.

It is important to forecast $\gamma_G^*$ and $\gamma_F^*$
to avoid waste of simulation.
The mean-field estimate is helpful for this purpose.
At the quenched level, the mean-field value provides a good
estimate of the anisotropy parameter of the gauge action.
In this work, we showed that the tuned anisotropy $\gamma_F^*$
approaches to the mean-field value from below as $\beta$
increases.
These may help us to select the values of $\gamma_G$ and
$\gamma_F$ to be explored.

Our preliminary result of the calibration of dynamical anisotropic
lattices with the same combination of actions as in this work
indicates that
the renormalized anisotropy defined through the gauge observable
less depends on $\gamma_F$, and one through the fermionic observable
less depends on $\gamma_G$ \cite{Nomura03}.
These dependences of $\xi_G$ and $\xi_F$ on
$\gamma_G$ and $\gamma_F$ indicate that linear fits may suffice
to determine $\gamma_G^*$ and $\gamma_F^*$  satisfying
$\xi_G(\gamma_G^*,\gamma_F^*) = \xi_F(\gamma_G^*,\gamma_F^*) = \xi$.
We would be able to perform the calibration by generating
configurations at best at four (or three) sets of $(\gamma_G,\gamma_F)$.


\section*{Acknowledgment}

We thank A. Nakamura, M. Okawa, and members of QCD-TARO collaboration
for useful discussions.
The numerical simulation was done on a NEC SX-5 at Research Center
for Nuclear Physics, Osaka University, and on a Hitachi SR8000
at High Energy Accelerator Research Organization (KEK).
H.M. and T.U. were supported by Japan Society for the Promotion of
Science for Young Scientists.


\end{document}